\documentclass[12pt]{article}
\usepackage{fullpage,doublespace}
\usepackage{amssymb, amsthm, amsmath}
\usepackage{graphicx}
\usepackage[authoryear]{natbib}
\usepackage{bm}
\usepackage{lineno}
\usepackage{times}

\newcommand{\balpha}{ \mbox{\boldmath $\alpha$}}

\newcommand{\bgamma}{ \mbox{\boldmath $\gamma$}}

\newcommand{\bvarepsilon}{ \mbox{\boldmath $\varepsilon$}}

\newcommand{\bSigma}{ \mbox{\boldmath $\Sigma$}}
\newcommand{\bOmega}{ \mbox{\boldmath $\Omega$}}

\newcommand{\bA}{ \mbox{\bf A}}
\newcommand{\ba}{ \mbox{\bf a}}

\newcommand{\bX}{ \mbox{\bf X}}

\newcommand{\bY}{ \mbox{\bf Y}}

\newcommand{\bs}{ \mbox{\bf s}}

\newcommand{\bu}{ \mbox{\bf u}}
\newcommand{\bv}{ \mbox{\bf v}}
\newcommand{\bV}{ \mbox{\bf V}}

\newcommand{\bC}{ \mbox{\bf C}}

\newcommand{\bH}{ \mbox{\bf H}}
\newcommand{\bI}{ \mbox{\bf I}}

\newcommand{\bM}{ \mbox{\bf M}}
\newcommand{\bW}{ \mbox{\bf W}}

\newcommand{\bQ}{ \mbox{\bf Q}}
\newcommand{\bU}{ \mbox{\bf U}}

\newcommand{\bzero}{ \mbox{\bf 0}}

\newcommand{\iid}{\stackrel{iid}{\sim}}
\newcommand{\indep}{\stackrel{indep}{\sim}}

\newcommand{\calR}{{\cal R}}

\newcommand{\calD}{{\cal D}}

\newcommand{\calA}{{\cal A}}

\newcommand{\Matern}{ \mbox{Mat$\acute{\mbox{e}}$rn}}

\newcommand{\beq}{ \begin{equation}}
\newcommand{\eeq}{ \end{equation}}
\newcommand{\beqn}{ \begin{eqnarray}}
\newcommand{\eeqn}{ \end{eqnarray}}

\newtheorem{assumption}{Assumption}

\begin{document}  

\thispagestyle{empty}

\begin{center}
{\Large A review of spatial causal inference methods for environmental and epidemiological applications}\\\vspace{6pt}
{\large Brian J Reich\footnote{North Carolina State University}, Shu Yang$^1$, Yawen Guan\footnote{University of Nebraska - Lincoln}, Andrew B Giffin$^1$, Matthew J Miller$^1$ and Ana G Rappold\footnote{US Environmental Protection Agency}}\\
\today
\end{center}

\begin{abstract}\begin{singlespace}
\noindent The scientific rigor and computational methods of causal inference have had great impacts on many disciplines, but have only recently begun to take hold in spatial applications.  Spatial casual inference poses analytic challenges due to complex correlation structures and interference between the treatment at one location and the outcomes at others.  In this paper, we review the current literature on spatial causal inference and identify areas of future work.  We first discuss methods that exploit spatial structure to account for unmeasured confounding variables.  We then discuss causal analysis in the presence of spatial interference including several common assumptions used to reduce the complexity of the interference patterns under consideration.  These methods are extended to the spatiotemporal case where we compare and contrast the potential outcomes framework with Granger causality, and to geostatistical analyses involving spatial random fields of treatments and responses.  The methods are introduced in the context of observational environmental and epidemiological studies, and are compared using both a simulation study and analysis of the effect of ambient air pollution on COVID-19 mortality rate.  Code to implement many of the methods using the popular Bayesian software {\tt OpenBUGS} is provided. \vspace{12pt} \end{singlespace}\end{abstract}

\newpage

\pagenumbering{arabic}
\section{Introduction}\label{s:intro}

Large-scale environmental and epidemiological studies often use spatially-referenced data to examine the effect of treatments or exposures on a health endpoint.  Examples include studying the effect of interventions on the spread of an infectious disease, pesticide application on cancer rates, and lead exposure on childhood development.  While standard analyses of spatial data simply estimate correlations, the ultimate goal of this research is to establish causal relationships \citep[e.g.,][]{bind2019causal} to inform decision making.  Therefore, developing statistical methods to establish causal relationships when data show spatial and temporal variation is invaluable to environmental science and epidemiology.

A rich literature on the theory and methods for causal inference for independent data has emerged \citep{bind2019causal,hernan2019causal}, but progress for spatial applications has been slow due to several analytic challenges.  First, randomization is often infeasible due to logistical or ethical concerns and so studies rely on observational data.  Second, exposure and response variables exhibit spatial correlation complicating statistical modeling and computation.  Third, the treatment at one location may influence the outcomes at nearby locations, a phenomenon known as spillover or interference.  These features of spatial applications violate the assumptions of standard causal inference methods and require new theory and computational tools.

Despite these challenges, major advances in spatial causal inference have been made in recent years.  In this paper, we review the recent progress on spatial causal inference, evaluate and compare current methods, and suggest areas of future work. We first review methods to adjust for missing spatial confounding variables \citep{hodges2010adding}. Most causal inference methods for observational data rely on an assumption of no missing confounding variables (i.e., unmeasured variables correlated with both the treatment and response).  However, if the missing confounding variables have prominent spatial patterns, methods have been developed to mitigate the bias caused by their omission.  These methods include case-control matching \citep[e.g.,][]{jarner2002estimation}, neighborhood adjustments by spatial smoothing \citep[e.g.,][]{schnell2019mitigating} and propensity-score methods \citep[e.g.][]{davis2019addressing}.  We review these methods and conduct a simulation study to compare their precision for estimating a causal treatment effect in the presence of a missing spatial  confounding variable.  A subset of the methods are applied to a study of the effect of ambient air pollution on the COVID-19 mortality rate.

A second major challenge in spatial causal inference is interference, where the treatment applied at one location affects the outcomes at other locations.  For example, an intervention to reduce the emissions from a power plant would affect the air quality at the power plant, but also locations downwind.  Capturing these spillover effects requires new definitions of the estimands of interest and new spatial models for the causal effects.  In full generality, allowing the treatment at a site to affect the outcomes at all other sites results in an intractable estimation problem. Therefore, assumptions are required to limit the form and spatial extent of interference.  We review several models for spatial interference including partial \citep[e.g.][]{zigler2012estimating} and network \citep[e.g.][]{tchetgen2017auto} interference.  We also discuss recent methods that combine mechanistic and spatial statistical models to anchor the causal analysis to scientific theory.

We begin reviewing these methods using cross-sectional data at a single time point, and then extend these methods to the spatiotemporal data. We discuss adapting spatial methods to the spatiotemporal setting, and methods specific to the temporal case such as difference-in-difference methods \citep[e.g.][]{delgado2015difference} that exploit changes over time to estimate causal effects. We also compare and contrast causal methods based on the potential outcomes framework \citep{rubin1974estimating} with Granger causality \citep{granger1969investigating}, which is defined specifically for processes that evolve over time.  We also discuss extensions of spatial methods for areal data defined at a finite number of regions (e.g., geopolitical units) to point-referenced (geostatistical) data in which case the treatment and response variables can be modeled as continuous random fields over an uncountable number of spatial locations.  This requires new definitions of causal effects, new methods for matching observations for case-control studies, and new models for missing spatial confounding variables and spillover effects. The paper concludes with a summary of the current literature and discussion of open problems in this rapidly-advancing field.


\section{Adjusting for spatial confounders}\label{s:confound}

To ensure privacy, public health data are often made available only after aggregation to administrative or geopolitical regions.  For areal data of this nature, we adapt the notation that $Y_{ij}$, $A_{ij}$ and $\bX_{ij} = (X_{ij1},...,X_{ijp})$ are the response, treatment and potential confounding variables (with $X_{ij1}=1$ for the intercept) for observation $j\in\{1,...,n_i\}$ in region $i\in\{1,...,N\}$ for a total of $n=\sum_{i=1}^Nn_i$ observations. The confounding variable in $\bX_{ij}$ can include both  covariates specific to observation $j$ within region $i$ or summaries of the region $i$ common to all $n_i$ observations in the region.   In addition to these observed variables, we allow for an unobserved confounding variable $U_i$ in region $i$, which is assumed to be a purely spatial term and thus the same for all observations in a region. 

{\bf Example 1}: As a concrete example, consider an environmental epidemiology study where $Y_{ij}$ is the birth weight of the $j^{th}$ baby born in zip code $i$ and $A_{ij}=1$ if the average ambient air pollution concentration in the mother's zip code exceeds a high threshold and $A_{ij}=0$ otherwise.  We may adjust for known confounding variables by including the mother's age and family income in $\bX_{ij}$, and describe the mother's environment by including the median income and measurable environmental factors such the average concentration of other known pollutants in region $i$ in $\bX_{ij}$.  In this scenario, the missing spatial confounder variable $U_i$ might be a second pollutant unknown to the researchers. The second pollutant qualifies as a missing spatial confounder if it has a strong spatial pattern, is associated with low birth weight, and is correlated with the pollutant of interest, perhaps via a common source.   Failing to account for this missing spatial confounder, either because its importance is unknown or data are unavailable, may inadvertently attribute the effects of the unknown pollutant to the pollutant of interest, biasing the estimator.

In this section we review spatial models for unknown processes such as $\bU=(U_1,...,U_N)^T$ (Section \ref{s:confound:review}) and causal inference methods that would apply if $\bU$ were observed (Section \ref{s:confounders:areal:PO}).  The remainder of the section is dedicated to methods that attempt to control for the missing confounder variable by exploiting its spatial structure.

\subsection{Review of spatial confounding}\label{s:confound:review}

Consider the spatial regression model
\begin{equation}\label{e:Y:areal}
Y_{ij} = A_{ij}\beta + \bX_{ij}\bgamma + U_i + \varepsilon_{ij},
\end{equation}
where $\beta$ is the treatment effect of interest, $\bgamma$ determines the effects of the confounding variables, $U_i$ is the spatial random effect for region $i$ and $\varepsilon_{ij}\iid\mbox{Normal}(0,\tau^2)$.  A common approach \citep{banerjee2014hierarchical} for areal data is to model the unobserved spatial effects  using a conditionally autoregressive (CAR) model (also known as a Gaussian Markov random field model).  The CAR model specifies spatial dependence in terms of the adjacencies between the regions.   The full conditional distribution of the random effect for one region given all other random effects is $U_i|U_k, k\ne i\sim\mbox{Normal}(\rho {\bar U}_i,\sigma^2/m_i)$, where ${\bar U}_i$ is the mean of $\bU$ at the $m_i$ regions adjacent to region $i$, and $\rho\in(0,1)$ and $\sigma>0$ are spatial covariance parameters.  These full conditional distributions define a multivariate normal distribution (Appendix A.2) for $\bU$,
which we denote as $\bU\sim\mbox{CAR}(\rho,\sigma)$.

The spatial regression model in (\ref{e:Y:areal}) where $\bU$ is modelled as a spatial process often gives very different estimates of covariate effects than the non-spatial model that excludes $\bU$, especially when the treatment variable exhibits a strong spatial pattern \citep{reich2006effects, paciorek2010importance, hodges2010adding}.  However, simply accounting for spatial correlation  does not resolve spatial confounding.  For example, Appendix A.1 describes a scenario where the bias of the posterior-mean estimator for $\beta$ depends on the strength of dependence between the treatment variable and the unmeasured confounding variable, but is the same whether the residuals are assumed to be independent or spatially correlated.  
The bias of this approach is confirmed in our simulation study (Section \ref{s:confounders:areal:sim}) when data are generated with correlation between $\bU$ and the treatment and response variables. This calls for methods that explicitly adjust for missing spatial confounders by blocking the dependence of $\bU$ on either the treatment or response variable.

\subsection{Potential outcomes framework}\label{s:confounders:areal:PO}

In this section we temporarily assume that $U_i$ is observed (and thus treated the same way as $\bX_{ij}$) to facilitate a review of standard non-spatial causal inference methods.  We begin with the potential outcomes framework \citep{rubin1974estimating}. Assume that the treatment $A_{ij}$ is binary and that each unit has two potential outcomes, $Y_{ij}(0)$ and $Y_{ij}(1)$, which represent the outcomes if the unit $j$ in region $i$ is given treatment $A_{ij}=0$ or $A_{ij}=1$, respectively. Our goal is to estimate the average treatment effect (ATE), \begin{equation}\label{e:ATE}
 \delta=\mbox{E}\left[\frac{1}{n}\sum_{i=1}^{N}\sum_{j=1}^{n_{i}}\{Y_{ij}(1)-Y_{ij}(0)\}\right],
\end{equation}
where the expectation is taken with respect to both $\bX_{ij}$ and $\{Y_{ij}(0),Y_{ij}(1)\}$. The fundamental problem is that only one of the two potential outcomes can be observed  \citep{holland1986statistics} rendering the other as counterfactual. Therefore, assumptions are required to ensure the ATE can be identified.

This notion of potential outcomes implicitly encodes the Stable Unit Treatment Values Assumption (SUTVA; \citealp{rubin1978bayesian}).  
\begin{assumption}[SUTVA]\label{asump:SUTVA} There is no interference and a single version of treatment.
\end{assumption}
\noindent SUTVA is violated under interference where $Y_{ij}$ depends not only on $A_{ij}$, but also on the treatment of other units. For instance, the birth weight of a baby in Example 1 could be influenced by the air pollution concentration both in the mother's zip code ($A_{ij}$) but also in other zip codes that the mother frequents. In this case, the potential outcomes are not determined by $A_{ij}$ alone, and we would need to introduce a different potential outcome for each combination of the treatment variables in the mother's vicinity (see Section \ref{s:spillover}).  

An example of multiple versions of treatment might be if birth weight actually depends not only on whether the air pollution exceeds a high threshold, but also a second extremely high threshold.  In this case, $A_{ij}$ actually has three levels (low, high and extremely high) and there should be three potential outcomes. An analysis that collapses the two high categories into a single group with $A_{ij}=1$ would violate SUTVA by having multiple versions of the treatment.  Violation of this assumption could be rectified by assuming $A_{ij}$ has three categories and thus each unit has three potential outcomes.

While SUTVA links treatments to potential outcomes, the consistency assumption is needed to further link the potential outcomes to the observations.   
\begin{assumption}[Consistency]\label{asump:consistency} The observed response is the potential outcome determined by the observed treatment variable, $Y_{ij} = Y_{ij}(A_{ij})$.
\end{assumption}

In addition to these assumptions about the treatment and response variables, a standard assumption that permits unbiased estimation of the ATE is the no missing confounder variables other than the observed covariates $X_{ij}$ and the latent spatial confounder $U_i$. We term this assumption as the latent ignorability assumption:
\begin{assumption}[Latent ignorability]\label{asump:LatentIgnorability}
	The potential outcomes $\{Y_{ij}(0),Y_{ij}(1)\}$ and treatments $A_{ij}$ are independent given $\bX_{ij}$ and $U_{i}$.
\end{assumption} 
\noindent Since $\bU$ is generally a latent (i.e., unknown) variable in the spatial setting, this assumption presumes that there exists some variable $\bU$ that blocks dependence between the treatment variable and potential outcomes; if $\bU$ is observed then this is the usual assumption that there are no unmeasured confounding variables.  This assumption implies that the confounding variables $\{\bX_{ij},U_i\}$ are sufficient to adjust for correlation between the observed treatment and response that is due to non-randomized treatment allocation and not an actual causal effect.  This requirement highlights the importance of careful evaluation of the system under study to ensure that all relevant variables are considered in $\bX_{ij}$.  

The final assumption deals with the distribution of observed treatment variables, i.e., the propensity score.  The propensity score is the probability of the treatment assignments, $\mbox{Prob}\{A_{ij}=1\mid \bX_{ij},U_{i},Y_{ij}(0), Y_{ij}(1)\}$.  Under Assumption \ref{asump:LatentIgnorability}, the propensity score becomes \begin{equation}\label{e:prop_score_def}
e(\bX_{ij},U_{i}) = \mbox{Prob}(A_{ij}=1\mid \bX_{ij},U_{i}).
\end{equation}
Assumption 4 is the standard positivity assumption on the propensity score:
\begin{assumption}[Positivity]\label{asump:positivity}
Both $e(\bX_{ij},U_{i})$ and $1-e(\bX_{ij},U_{i})$ are positive for all $\bX_{ij}$ and $U_i$.
\end{assumption} 
\noindent This assumption implies that both $A_{ij}=0$ and $A_{ij}=1$ are possible under the treatment allocation mechanism, which is necessary to estimate the ATE in (\ref{e:ATE}) which averages over the expected potential outcome under both treatments.  

Under Assumption \ref{asump:LatentIgnorability} the propensity score is a function of known variables $\bX_{ij}$ and $U_i$ and can thus be estimated without knowledge of unobservable counterfactual responses.  However, Assumptions 1-3 are difficult or impossible to verify empirically, and thus a causal inference requires scrutinizing the study design and the processes of interest to justify that these assumptions hold.  One of the main contributions of causal inference is to state explicitly the assumptions needed for an estimator to have a casual interpretation, and thus guide a discussion of a study's results.

Assumptions 1--4 underlie many non-spatial causal estimation procedures such as (augmented) inverse probability weighting \citep[e.g.,][]{rosenbaum&rubin83a,robins1994adjusting,bang2005doubly,cao2009improving}, and matching \citep[e.g.,][]{rosenbaum1989optimal,heckman1997matching,hirano2003efficient,hansen2004full,rubin2006matched,abadie2006large,stuart2010matching,abadie2016matching}.  To fix ideas, we focus on the simplest approach of the linear model in (\ref{e:Y:areal}) where $U_i$ is observed and thus not given a spatial model.
Spatial analyses often rely on parametric models because the lack independent replications in a region complicates non-parametric methods.  The parametric model in (\ref{e:Y:areal}) makes the additional assumptions of linearity and normality, but gives valid causal inference under the assumed model and Assumptions 1-4. In other words, the regression coefficient $\beta$ can be interpreted as the ATE, $\delta$.  Therefore, if $U_i$ is observed and these assumptions hold, then the estimate of $\beta$ from a standard least squares analysis has a causal interpretation.  In the remainder of this section we discuss methods to deal with unknown $\bU$.

\subsection{Case-control matching methods}\label{s:confounders:areal:matching}

While most of the methods we discuss control for confounding at the analysis stage, a case-control study controls for confounding at the design stage.  In a case-control analysis of a binary response variable (i.e., $Y_{ij}\in\{0,1\}$), each case ($Y_{ij}=1$) is matched with one or more controls ($Y_{ij}=0$) that are drawn from the same underlying population at risk. When applying this study design, investigators sample controls to resemble cases with respect to all factors that may determine the disease status except for the exposure of interest. As discussed below, this design removes the need to adjust for the matching factors at the analysis stage.  Matching variables can be specific to the individual, such as age or education level.  Partial control for spatial variation of risk can be achieved by matching on confounding factors that vary spatially such as the region's median income.  To adjust for unmeasured spatial confounders, controls can be matched based on their proximity to the cases \citep{jarner2002estimation}. Assuming there is replication within region $(n_i>1)$ and treatment varies within region $(A_{ij}\ne A_{il}$ for some $j$ and $l$) then matching individuals in the same region is an effective means of adjusting for spatial confounding.

Matched case-control data are most often analyzed using conditional logistic regression. Assume each case $Y_{ij}=1$ is paired with a single control $Y_{kl}=0$.  Under the spatial logistic regression model $\mbox{logit}\{\mbox{Prob}(Y_{ij}=1)\} = A_{ij}\beta+\bX_{ij}\bgamma+U_i$,  the log odds that $Y_{ij}=1$ given either $Y_{ij}=1$ or $Y_{kl}=1$ (but not both) is $$\eta_{ij}=  (A_{ij}-A_{kl})\beta+(\bX_{ij}-\bX_{kl})\bgamma + U_i-U_{k}.$$   
To account for variability within each pair (strata), a random intercept $z_{ij}$ is added so the likelihood contribution of the pair is $$\mbox{Prob}(Y_{ij}=1|Y_{ij}=1+Y_{kl}=1)=\exp(\eta_{ij}+z_{ij})/\{1+\exp(\eta_{ij}+z_{ij})\}.$$  Since the covariates appear in the likelihood only through the difference $\bX_{ij}-\bX_{kl}$, the effect of covariates used for matching cannot be estimated and these covariates can be removed from the model. Similarly, if cases are paired with observations from the same region (i.e., $i=k$), then the spatial random effects $\bU$ do not appear in the likelihood and a non-spatial analysis is sufficient. Thus, while the matched case-control analysis is an excellent means of controlling for confounders, its drawbacks include discarding data and not being able to estimate all covariate effects and spatial variation in risk.

Pairing observations in the same region can also be applied for continuous responses.  For a continuous response there is no natural definition of a case or control, but regressing the difference between the responses in the same region removes spatial confounding.  For example, under the linear model in (\ref{e:Y:areal}) the model for the difference between responses in the same region is
\begin{equation}\label{e:matching_continuous}
Y_{ij}-Y_{il}=(A_{ij}-A_{il})\beta+(\bX_{ij}-\bX_{il})\bgamma+{\tilde \epsilon}_{i},\end{equation}
where ${\tilde \epsilon}_{i}$ is independent error.  Again, differencing eliminates the latent variable $U_i$, and thus the differences can be analyzed with non-spatial methods.  This approach relies on a parametric linear model, but the concept of reducing bias by pairing observations in the same location can also be applied using weighting based on the propensity score model \citep{he2018inverse}.

\subsection{Neighborhood adjustments}\label{s:confounders:areal:neighborhood}
In (\ref{e:matching_continuous}),  modelling the difference between observations in the same region eliminated the unmeasured confounders.  In cases without replication and a missing confounder that varies smoothly across space, its effect can be reduced by removing large-scale spatial trends from the response, the treatment, or both.  Removing large-scale trends isolates local variation in the response, which is arguably less prone to spatial confounding than large-scale variation.  In this section we review several methods that have been proposed for removing large-trends in spatial regression.

\subsubsection{Simultaneous Autoregressive models}

For simplicity, assume there are no replications within each region and temporarily drop the replication subscript by defining $Y_{i1}=Y_i$, $\bX_{i1}=\bX_i$ and $A_{i1}=A_i$.  Rather than specifying the regression on the response, the Simultaneous Autoregressive (SAR) model first subtracts regional means
\begin{equation}\label{e:SAR:uni}
 Y_i - \phi{\bar Y}_i = (A_i-\phi{\bar A}_i)\beta + (\bX_i-\phi{\bar \bX}_i)\bgamma + \varepsilon_i,
\end{equation}
where ${\bar Y}_i$, ${\bar A}_i$ and ${\bar \bX}_i$ are the means of the response, treatment and covariates at the $m_i$ regions adjacent to region $i$, $\phi$ is an unknown parameter and $\varepsilon_i\iid\mbox{Normal}(0,\sigma^2)$.  Taking differences reduces the effect of missing confounding variables that are constant across neighboring regions.  In vector form, (\ref{e:SAR:uni}) can be expressed as $\bY = \bA\beta + \bX\bgamma + \bvarepsilon$ where the spatial covariance of $\bvarepsilon$ is given in Appendix A.2. \cite{wall2004close} compares differences in covariance implied by the SAR and CAR models.  \cite{wall2004close} finds the models produce similar regression coefficient estimates despite sometimes large differences in covariances between regions.

\subsubsection{Neighborhood adjustment via spatial smoothing}
Rather than simply subtracting the mean of neighboring sites, spatial trends can be removed by joint spatial modeling of the treatment and the missing spatial confounder. Consider the spatial regression model in \eqref{e:Y:areal} without replicates. The bias is a result of attributing the effect of the confounder on $\bY$ to the treatment variable when $\bA$ and $\bU$ are correlated (Appendix 1). \cite{schnell2019mitigating} provide a set of assumptions  (given in the Appendix) to identify the unmeasured confounding bias $\mbox{E}(U_i|\bA) = B_i(\bA)$. They model $B_i(\bA)$ by specifying a joint distribution for $\bU$ and $\bA$ that allows each process to have a different range of spatial correlation and permits correlation between $\bU$ and $\bA$.  The confounding bias is mitigated by fitting a joint model \begin{eqnarray}\label{e:CAR:adjust1}
Y_i &=& A_i\beta - B_i(\bA) + \bX_i\bgamma + {\bf e}_{i1} \\\label{e:CAR:adjust2}
A_i &=& \bX_i\balpha + {\bf e}_{i2},\nonumber
\end{eqnarray}
where the form of $B_i(\bA)$ and the spatial covariance of $e_{i1}$ and $e_{i2}$ are given in Appendix A.3.  As noted by \cite{schnell2019mitigating} and was also suggested by \cite{paciorek2010importance}, if the spatial scale of treatment is larger or about the same as the unmeasured confounder, the confounding bias cannot be mitigated. 

\subsection{Propensity score methods}\label{s:confounders:areal:propscores}

Propensity scores are used in a wide range of causal inference methods.  Assuming a binary treatment variable, the propensity score for observation $j$ in region $i$ is $\mbox{Prob}(A_{ij}=1)=e_{ij}$.  In a standard analysis the propensity scores are modeled as a function of the known covariates $\bX_{ij}$ and the estimated propensity scores are used to alleviate the imbalance of the covariates between treatment groups.  Here we face the additional challenge that the propensity scores may depend on the unobserved spatial process, $U_i$.  

For example, consider the simple hierarchical model that includes the unobserved spatial process in the propensity score,
\begin{eqnarray}\label{e:areal:jointY}
Y_{ij} &=& A_{ij}\beta + \bX_{ij}\bgamma + U_i +\varepsilon_{ij}\\
 A_{ij}&\sim&\mbox{Bernoulli}(e_{ij})\mbox{\ \ \ \ \ with \ \ \ \ \ }\mbox{logit}(e_{ij}) = \bX_{ij}\balpha  + \phi U_i + V_i,\label{e:areal:jointA}
\end{eqnarray}
where $V_i$ accounts for spatial patterns in treatment allocation not accounted for by the covariates or the missing confounder $U_i$.  To emphasize the effect of the propensity score on the response model, (\ref{e:areal:jointY})-(\ref{e:areal:jointA}) can be reparameterized ($U_i=u_i+\gamma v_i$ and $V_i=v_i-\psi u_i+\phi\psi v_i$) as
\begin{eqnarray}\label{e:areal:joint2Y}
Y_{ij} &=& A_{ij}\beta + \bX_{ij}\bgamma + u_i +\psi v_{i}+\varepsilon_{ij}\\
 A_{ij}&\sim&\mbox{Bernoulli}(e_{ij})\mbox{\ \ \ \ \ with \ \ \ \ \ }\mbox{logit}(e_{ij}) = \bX_{ij}\balpha  + v_i.\label{e:areal:joint2A}
\end{eqnarray}
The shared spatial random effect $v_i$ adjusts for the missing confounder by absorbing signal in the response that can be explain by spatial trends in the treatment allocation.  The spatial random effects can be assigned priors $\bu=(u_1,...,u_N)^T\sim\mbox{CAR}(\rho_u,\sigma_u)$ independent of $\bv=(v_1,...,v_N)^T\sim\mbox{CAR}(\rho_v,\sigma_v)$.   Fitting this joint model for the treatment and response processes is straightforward using hierarchical Bayesian methods.

A concern with this model is that some of its many parametric assumptions could be violated, invalidating inference.  Another issue is that of so-called ``feedback'', which in this context refers to information in the response influencing the posterior of the  propensity scores \cite[e.g.,][]{zigler2013model,zigler2016central,saarela2016bayesian}.  Eliminating this feedback can be done by fitting the model in two stages, i.e., first fitting the model for the treatment indicators in (\ref{e:areal:joint2A}) to obtain an estimate of $\bv$ and then fitting (\ref{e:areal:joint2Y}) with $\bv$ fixed at its first-stage estimate.  Other possible remedies include ``cutting feedback'' in the steps of the MCMC algorithm \citep{lunn2009combining, mccandless2010cutting} or post-hoc reweighting of the posterior distribution \citep{saarela2015bayesian,davis2019addressing}.  These methods are discussed below.

Referring to the joint model in (\ref{e:areal:joint2Y})-(\ref{e:areal:joint2A}), if the propensity score $e_{ij}$ were known and $\mbox{logit}(e_{ij})$ were included as a known confounder in $\bX_{ij}$, then latent ignorability (Assumption \ref{asump:LatentIgnorability}) would hold and the resulting estimate of $\beta$ would have a causal interpretation.  Of course, the exact propensity is unknown and must be estimated. Let ${\hat e}_{ij}$ be a first-stage propensity-score estimator, e.g., as estimated by fitting the spatial logistic regression model in (\ref{e:areal:joint2A}).  The estimated propensity scores can be included in the mean of the response model to account for spatial confounding.   The propensity score can be added to the response model as,
\begin{equation}\label{e:Y:areal:propadj}
Y_{ij} = A_{ij}\beta + \bX_{ij}\tilde \bgamma + U_i + f({\hat e}_{ij})+\varepsilon_{ij}
\end{equation}
where $f$ is the logit function or more generally a non-linear function estimated by, say, smoothing splines. Given the inclusion of the propensity score, it can now be assumed that $U_i$ and $A_{ij}$ are conditionally independent. Assuming the model assumptions hold and the propensity score estimate is accurate, then $\beta$ has a causal interpretation.

Alternatively, the propensity score estimates can be used to define strata, i.e., 
\begin{equation}\label{e:Y:areal:propadj2}
Y_{ij}|{\hat e}_{ij}\in [T_l,T_{l+1}) = S_l + A_{ij}\beta + \bX_{ij}\bgamma + U_i +\varepsilon_{ij}
\end{equation}
where $0=T_1<T_2<....<T_{L+1}=1$ define the propensity-score strata, $S_l$ encodes the unmeasured confounder effect for strata $l$ and $U_i$ and $A_{ij}$ are conditionally independent.  Although the strata are defined irrespective of spatial information, the spatial random effect $U_i$ accounts for spatial dependence.

This joint modeling framework can be extended to continuous treatment variables by replacing the the Bernoulli/logistic model for $A_{ij}$ in (\ref{e:areal:joint2A}) with a normal model with $\mbox{E}(A_{ij}\mid \bX_{ij},v_i)=e_{ij}=\bX_{ij}\balpha  + v_i$ and  $\mbox{Var}(A_{ij}\mid \bX_{ij},v_i)=\sigma_A^2$.  This method could be fit as a joint model or in two stages where first a Gaussian spatial model for $A_{ij}$ is fit and estimates of $e_{ij}$ are used as generalized propensity scores \citep{hirano2004propensity} in the response model as in (\ref{e:Y:areal:propadj}) or (\ref{e:Y:areal:propadj2}).  Generally, this model-based framework can be adapted to more complex settings as long as a model with reasonable fidelity to the data generating process can be determined and justified.  

As an alternative to model-based causal adjustment, \cite{davis2019addressing} use imputation of potential outcomes and propensity-score weighting. They first estimate propensity scores ${\hat e}_{ij}$ using a spatial regression such as (\ref{e:areal:joint2A}).  Then in a second stage, they fit the response model in (\ref{e:Y:areal}), which excludes the propensity score. Rather than use the estimate of $\beta$ from this analysis, they post-process the model output to remove confounding bias.  They estimate the causal effect using concepts from augmented inverse probability weighting \citep{rosenbaum1983assessing,robins1994estimation, bang2005doubly, cao2009improving}
\begin{eqnarray}\label{e:weighting}
    \delta &=& \frac{1}{N}\sum_{i=1}^N\sum_{j=1}^{n_i}\delta_{ij}\\  
    \delta_{ij} &=& \frac{1}{{\hat e}_{ij}}\left\{A_{ij}Y_{ij}-(A_{ij}-{\hat e}_{ij}){\tilde Y}_{ij1}\right\} -  
    \frac{1}{1-{\hat e}_{ij}}\left\{(1-A_{ij})Y_{ij}-({\hat e}_{ij}-A_{ij}){\tilde Y}_{ij0}\right\}\nonumber
  \end{eqnarray}
where ${\tilde Y}_{ija} = a{\hat \beta} + \bX_{ij}{\hat \bgamma} + {\hat U}_i$ is the estimated mean response setting $A_{ij}=a$ for $a\in\{0,1\}$.  \cite{davis2019addressing} suggest using bootstrap sampling (which account for uncertainty at all stages) or a closed form large-sample variance estimator to quantify uncertainty in $\delta$.  Alternatively, in a Bayesian analysis, samples from the posterior distribution of $\delta$ can be made by computing $\delta$ for each posterior sample of $\{\beta, \bgamma,\bU\}$.

\subsection{Instrumental variables}\label{s:confounders:areal:IM}

\begin{figure}
	\centering
	\caption{{\bf A directed acyclic graph (DAG) represents the dependence of the random variables.} $Z$ is the instrumental variable, $A$ is the treatment, $Y$ is the outcome, $X$ is the observed confounder, and $U$ is the unobserved confounder.}\label{fig:IV}		
	\includegraphics[width=.3\textwidth]{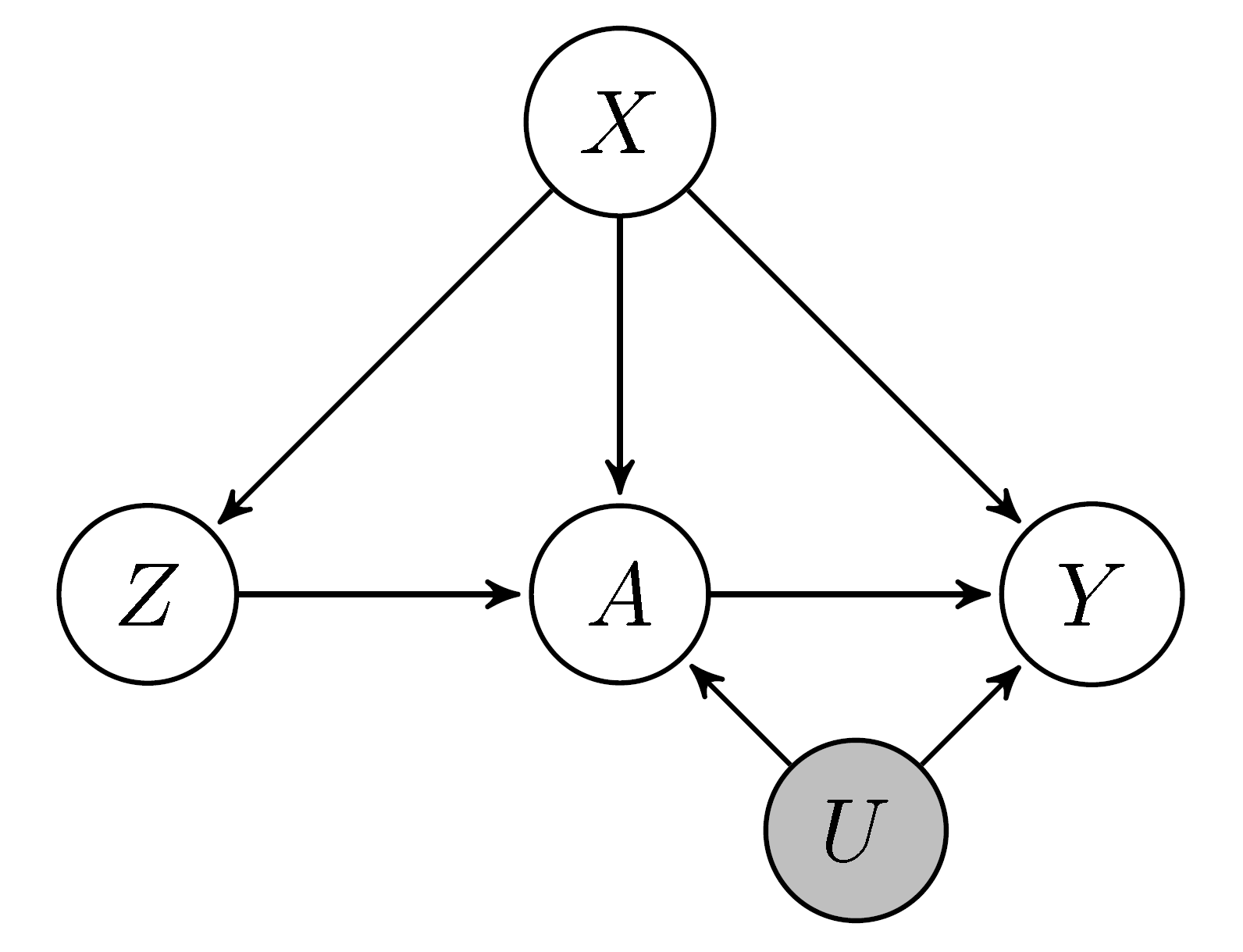}
\end{figure}

An instrumental variable (IV) $Z_i$ is widely used to deal with unmeasured confounding. An valid IV must (a) be associated with the treatment $A_i$, (b) not be related to the unmeasured confounder $U_i$, and (c) not be directly affect the outcome.  Figure \ref{fig:IV} illustrates the dependence structure of the random variables. As an example, suppose $A_i$ is a region's air pollution level and $Y_i$ is the region's asthma rate.  A potential instrumental variable is the region's traffic density, $Z_i$.  As traffic is a major source of air pollution, it is clear that $A_i$ and $Z_i$ are correlated, and it can be argued that traffic density is unrelated to asthma rate other than via air quality. 

The classic causal analysis with IVs is a two-stage least squares regression, The treatment is first regressed onto the IV, and then the fitted values from this first-stage regression as used as the treatment variable in the response model.  That is, if the first-stage regression gives ${\hat A}_i = {\hat \alpha}_0 + Z_i{\hat \alpha}_1 + \bX_{i}{\hat \alpha}_2 $, then the second stage model replaces $A_i$ with $Z_i{\hat \alpha}_1$, i.e., 
$Y_{i} = {\hat \alpha_1}Z_i\beta + \bX_{i}\bgamma  + \varepsilon_{i}$.  This confines the treatment variable to the span of the instrumental variable, and thus to a space orthogonal to the missing confounding variable.  If a valid IV can be identified then this provides a simpler means of estimating average treatment effect instead of adjusting for missing confounders than propensity scores.  

Some caution has to be exercised when interpreting causal estimates based on IVs. In the observational setting, as in traffic instrument example, the investigators do not have the ability to enforce treatment (PM) based on treatment assignment (traffic). Although traffic is a major source of variation in PM,  other sources can play a role which leads to differences between intended and observed treatments among units and potentially to the heterogeneity of responses  (power plants, wildfires, etc). In randomized treatment-control examples, this equates to the lack of full compliance between treatment assignment and the intake of drug. The implication is that the ATE is estimated only among those whose PM variation is explained by variation in the instrumental variable, referred to as the local average treatment effect (LATE) or complier average treatment effect (CATE).  \cite{Imbens1994Identification} provide the criteria under which the LATE/CATE represents the ATE. 

Spatial consideration can be made in both stages of the model.  Consider a continuous treatment variable and the joint model
 \begin{eqnarray}
Y_{ij} &=& \alpha_1Z_{ij}\beta + \bX_{ij}\bgamma + U_i +\epsilon_{1ij}\label{e:areal:IVY}\\
 A_{ij} &=& \alpha_0 + Z_{ij}\alpha_1 + \bX_{ij}\balpha+ \phi U_i+V_i+\epsilon_{2ij},\label{e:areal:IVA}
\end{eqnarray} 
where $\bU\sim\mbox{CAR}(\rho_U,\sigma_U)$, $\bV\sim\mbox{CAR}(\rho_{V},\sigma_{V})$\label{e:areal:IV},
$\epsilon_{1ij}\iid\mbox{Normal}(0,\tau^2_1)$ and $\epsilon_{2ij}\iid\mbox{Normal}(0,\tau^2_2)$. In (\ref{e:areal:IVY}), $A_{ij}$ in the response model in (\ref{e:Y:areal}) is replaced by  $Z_{ij}\alpha_1$ in the instrumental variable regression. Spatial random effects are included in both stages of the model to provide more efficient estimators of the regression coefficients and valid uncertainty quantification.  This model closely resembles the joint propensity score model in (\ref{e:areal:jointY})-(\ref{e:areal:jointA}) except that only the signal in $A_{ij}$ than can be explained by the IV enters the response model. 

The two models in (\ref{e:areal:IVY})-(\ref{e:areal:IVA}) can be fit simultaneously, although feedback effects must be considered as in the propensity score methods of Section \ref{s:confounders:areal:propscores}.  Alternatively, the method can be fit in two stages.  The first stage is a spatial regression of $A_i$ onto $Z_i$ in (\ref{e:areal:IVA}) and $\bX_i$ gives an estimate of $\alpha_1$.  In the second stage spatial regression of the response,  $Z_i{\hat \alpha}_1$ is used as the treatment variable.  An important difference between the classical and this spatial IV approach is that in the spatial version the fitted values will not be strictly orthogonal to the errors $U_i$. A potential remedy is the use of restricted spatial regression \citep{reich2006effects,hodges2010adding,hughes2013dimension,hanks2015restricted}. 

\subsection{Structural equation modeling}\label{s:confounders:SEM}
\cite{thaden2018structural} propose to adjust for spatial confounding using structural equation modelling (SEM).  They introduce binary indicator variables for each spatial location in both the models for the treatment and response variables.  Therefore, although motivated using SEMs, they arrive at a similar model to the joint model in  (\ref{e:areal:joint2Y})-(\ref{e:areal:joint2A}).  They argue that independent priors for the random effects ($u_i$ and $v_i$ in (\ref{e:areal:joint2Y})-(\ref{e:areal:joint2A})) more effectively resolve spatial confounding than spatial priors. Treating the random effects as independent requires replication within region, which is not always available.  However, when there is sufficient replication within regions, independent priors are preferable to spatial models because they are less constrained and thus more completely block spatial confounding. 

\subsection{Simulation study}\label{s:confounders:areal:sim}

In this section we conduct a simulation study to compare methods for adjusting for an unmeasured confounding variable.  We examine how the methods compare with different levels of spatial correlation in the treatment and confounding variable, and robustness to model misspecification.  

{\bf Data generation}: We generate data from the  model
\begin{equation}\label{e:sim:gen}
Y_i|A_i \indep \mbox{Normal}(A_i\beta + U_i,1) \mbox{\ \ \ \ \ and \ \ \ \ \ }
 A_i\indep\mbox{Bernoulli}\left[\mbox{expit}\left\{g(V_i,\phi U_i)\right\}\right]
\end{equation}
where the spatial terms are drawn from the model $\bU\sim\mbox{CAR}(\rho_U,2)$, $\bV\sim\mbox{CAR}(\rho_V,2)$ and the transformation function $g$ is given below.  The correlation structure is determined by three parameters: $\rho_U$ and $\rho_V$ control the range of spatial dependence and $\phi$ controls the strength of spatial confounding. For simplicity we exclude known confounders $\bX_i$ to isolate the effects of spatial confounding.  The first four scenarios have $g(V_i,\phi U_i)=V_i+\phi U_i$ and vary $\rho_U,\rho_V\in\{0.90,0.99\}$ to study the performance of the joint model when it is correctly specified. Setting the CAR dependence parameter to 0.99 gives strong spatial dependence with correlation 0.54 between adjacent regions in the center of the grid, while the value 0.90 gives moderate correlation of 0.35 between adjacent regions in the center of the grid. The final two scenarios have $\rho_U=\rho_V=0.99$ but nonlinear function $g(V_i,\phi U_i)=V_i+\phi \{U_iI(U_i>0)-0.63\}$ (``Nonlinear'') or the nonstationary function $g(V_i,\phi U_i)=V_i+\phi U_ic_i$ where $c_i$ increases linearly from zero to one across the columns of the grid (``Nonstationary'').  These scenarios are included to investigate the performance of the joint model when it is misspecified.

We generated 100 datasets on a $30\times30$ square grid of regions with rook neighbors and $\beta=\phi=0.5$.   For each dataset we fit the following models.
\begin{itemize}
\item {\bf NS}: Non-spatial least squares, $Y_i\indep\mbox{Normal}(\gamma+A_i\beta,\tau^2)$
\item {\bf NS+P}: Non-spatial least squares with a spline function of the propensity score,\\ $Y_i\indep\mbox{Normal}\left\{\gamma+A_i\beta+f({\hat e}_i),\tau^2\right\}$
\item {\bf S}: Spatial CAR regression without confounder adjustment, $Y_i\indep\mbox{Normal}(\gamma+A_i\beta+U_i,\tau^2)$
\item{\bf S+P}: Spatial CAR regression with a spline function of the spatial propensity score,\\$Y_i\indep\mbox{Normal}(\gamma+A_i\beta+U_i+f({\hat e}_i),\tau^2)$
\item {\bf S+AIPW}: Spatial CAR regression with post-hoc IDW debiasing step, i.e., model S with post-processing as in (\ref{e:weighting})
\item {\bf Joint}: Joint model in (\ref{e:areal:joint2Y})-(\ref{e:areal:joint2A})
\item {\bf Cut}: Joint model with feedback cut as in \cite{mccandless2010cutting}
\end{itemize}
In these models ${\hat e}_i $ is computed using the spatial logistic regression in (\ref{e:areal:joint2A}) and $f$ is a B-spline basis expansion with five degrees of freedom.  The priors for all models are $\bU\sim\mbox{CAR}(\rho_U,\sigma_U)$, $\bV\sim\mbox{CAR}(\rho_v,\sigma_v)$, $\rho_U,\rho_V\sim\mbox{Uniform}(0,1)$, all mean parameters have $\mbox{Normal}(0,10)$ priors and all variances have $\mbox{InvGamma}(0.5,0.005)$ priors.  All of these method are fit in {\tt OpenBUGS} and the code is available at {\tt https://github.com/reich-group/SpatialCausalReview/}.

\begin{figure}\caption{{\bf Simulation study results}. The boxplots summarize the sampling distribution of the causal estimates across datasets and the solid line at 0.5 is the true value.  The scenarios vary by the spatial dependence parameter of the confounder ($\rho_u$) and treatment ($\rho_v$) variables, and whether the joint model is misspecified.  The competing methods are defined in Section \ref{s:confounders:areal:sim}.   The empirical coverage of 95\% credible intervals for the causal effect are given above the model labels.  }\label{f:sim}
\centering 
\includegraphics[page=1,width=0.32\textwidth]{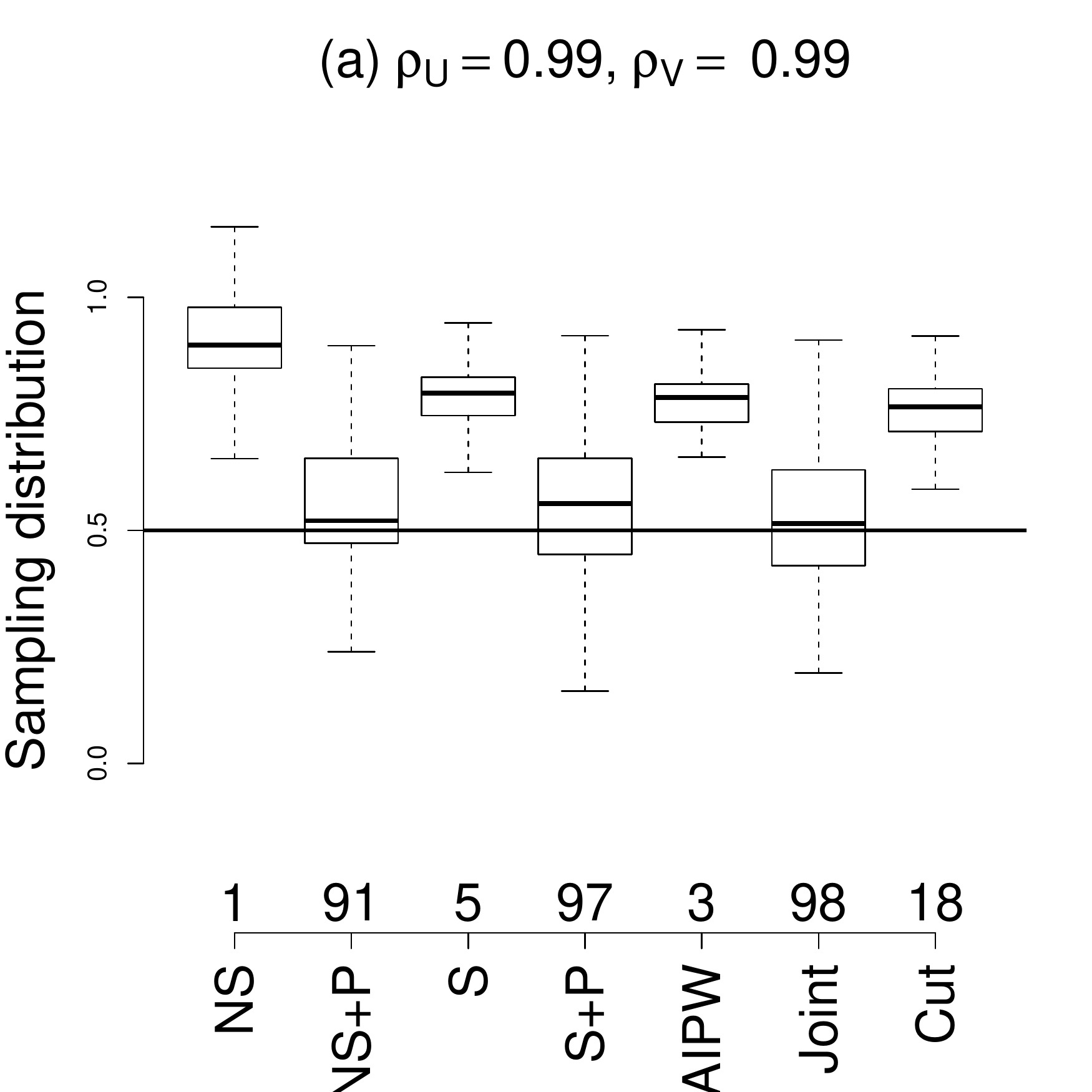}
\includegraphics[page=2,width=0.32\textwidth]{figs/Sim}
\includegraphics[page=3,width=0.32\textwidth]{figs/Sim}\vspace{12pt}
\includegraphics[page=4,width=0.32\textwidth]{figs/Sim}
\includegraphics[page=5,width=0.32\textwidth]{figs/Sim}
\includegraphics[page=6,width=0.32\textwidth]{figs/Sim}
\end{figure}

Figure \ref{f:sim} plots the causal effect estimates across datasets for each scenario and statistical method. As expected, the non-spatial method (NS) without causal adjustment is biased and has low coverage in all cases.  The spatial model without causal adjustment (S) provides only a small improvement.  The non-spatial model with spatial propensity score (NS+P) substantially reduces bias although its coverage remains below the nominal level.  The spatial model with causal post-processing (AIPW) and the joint model that cuts feedback (Cut) have large bias and low coverage in the cases we considered.  

In this simulation the most effective methods are the spatial model with propensity score adjustment (S+P) and the full joint model (Joint).  This is not surprising in the first four scenarios because the joint model was used to generate the data.  In these cases the joint model appears to have less bias than the two-stage spatial propensity score model, but both methods are similar.  These models are misspecified in the final two scenarios, but still outperform the other methods.  Surely more extreme scenarios where these methods fail to deliver reliable inference can be devised, but these results suggest some robustness to model assumptions.

The strength of the spatial correlation in the treatment allocation process appears to be more predictive of reliable performance than model misspecification.  In scenarios (b) and (d) with $\rho_U=0.9$ all of the methods are biased and have low coverage.  In these cases the spatial model of the treatment allocation process has low predictive power and thus all subsequent causal adjustments are ineffective.  In these cases the unmeasured confounder cannot be explained by known covariates or spatial patterns, and there is simply no structure that can be exploited to remove its effect.

\subsection{Effect of PM$_{2.5}$ exposure on COVID-19 mortality}\label{s:confounders:areal:COVID}

\begin{figure}\caption{{\bf Plots of the COVID-19/PM$_{2.5}$ data}. Panel (a) plots the sample log COVID-19 mortality rate, $\log(Y_i/N_i)$, through May 12, 2020 with gray denoting no observed deaths ($Y_i=0$); Panel (b) maps the long-term (2000-2016) average fine particulate matter (PM$_{2.5}$) concentration.  Alaska and Hawaii are excluded from the study.}\label{f:covid_maps}
\centering \vspace{-64pt}
\includegraphics[page=1,width=0.7\textwidth]{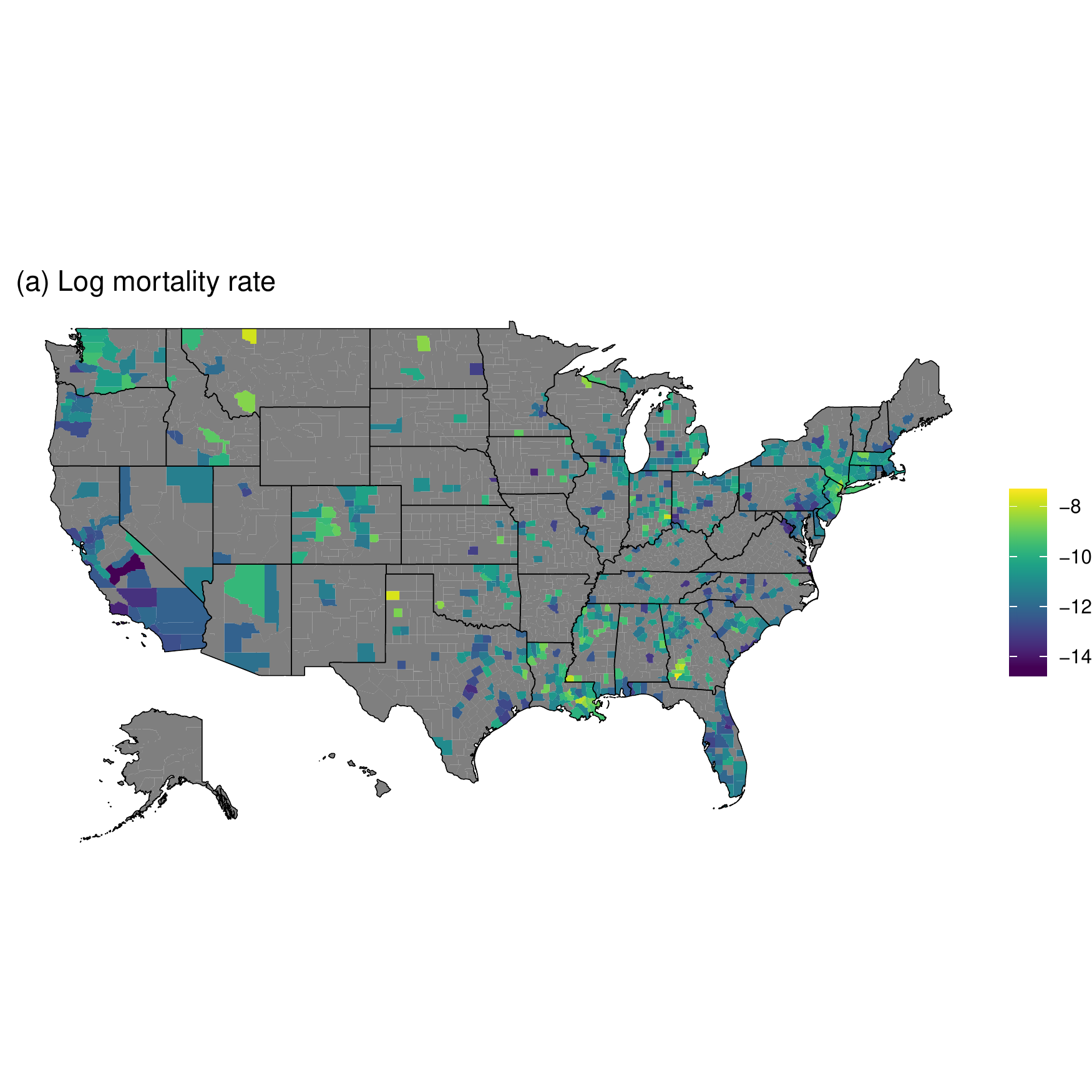}\\\vspace{-120pt}
\includegraphics[page=2,width=0.7\textwidth]{figs/COVID_PM.pdf}
\end{figure}

To illustrate the spatial confounder adjustment methods, we reanalyze the data provided by \cite{wu2020exposure}.  The response $Y_i$ for county $i$ is the number of COVID-19 related deaths through May 12, 2020.  The treatment variable $A_i$ is the long-term (2000-2016) average fine particulate matter (PM$_{2.5}$) concentration.  These variables are plotted in Figure \ref{f:covid_maps} and both show strong spatial trends.  The known confounder variables in $\bX_i$ include $p=15$ measures of the county's demographic, socio-economic and climate conditions (see Table 2 of \cite{wu2020exposure} for a complete list).  Some covariates (number of hospital beds, BMI and smoking rate) have a high proportion of missing values.  Rather than removing the counties with missing value, which would complicate the spatial adjacency structure, we remove the covariates with missing value.  Removing these covariates does not greatly affect the effect estimates (as discussed below).

Because the dataset is large and the treatment is continuous we consider only the non-spatial (``NS'') and spatial (``S'') models and these models with a two-stage propensity score adjustment (``NS+P'' and ``S+P'').   The response model is $Y_i\sim\mbox{Poisson}(N_i\lambda_i)$ where $N_i$ is the county's population and $\lambda_i$ is the mortality rate. \cite{wu2020exposure} use a quasi-Poisson model with state-level random effects; we use county-level random effects and allow these random effects to account for over-dispersion. Specifically, the mortality rate is modeled as
\begin{equation}\label{e:Y_covid}
\log{\lambda_i} = A_i\beta + \bX_i\bgamma + U_i + f({\hat e}_i)
\end{equation}
where $\bU\sim\mbox{CAR}(\rho_u,\sigma_u)$, ${\hat e}_i$ is the estimated generalized propensity score \citep{hirano2004propensity}, and $f$ is a B-spline basis with 5 degrees of freedom.     The generalized propensity score is the fitted negative log-likelihood (ignoring constants) ${\hat e}_i = (A_i-\bX_i{\hat \balpha}-{\hat V}_i)^2$, where ${\hat \balpha}$ and ${\hat V}_i$ are the posterior means from the model $A_i = \bX_i\balpha+V_i+\varepsilon_i$ and $\bV\sim\mbox{CAR}(\rho_v,\sigma_v)$ and $\varepsilon_i\iid\mbox{Normal}(0,\sigma^2_e)$.  The priors are the same as in Section \ref{s:confounders:areal:sim}.  The non-spatial models set $\rho_u=0$ (the county-level random effect remain in the model to account for overdispersion) and the methods without a propensity score set $f({\hat e}_i)=0$.

\begin{figure}\caption{{\bf Causal effect estimate for the COVID-19/PM$_{2.5}$ analysis}. Posterior distribution of the log relative risk of an increase of 1 $\mu g/m^3$ in long-term average PM$_{2.5}$ ($\beta$) on a county's COVID-19 mortality rate.  The four models are defined by whether they are non-spatial (``NS'') or spatial (``S'') and whether they include a spatial propensity score (``+ P'').  }\label{f:covid_beta}
\centering
\includegraphics[page=3,width=0.5\textwidth]{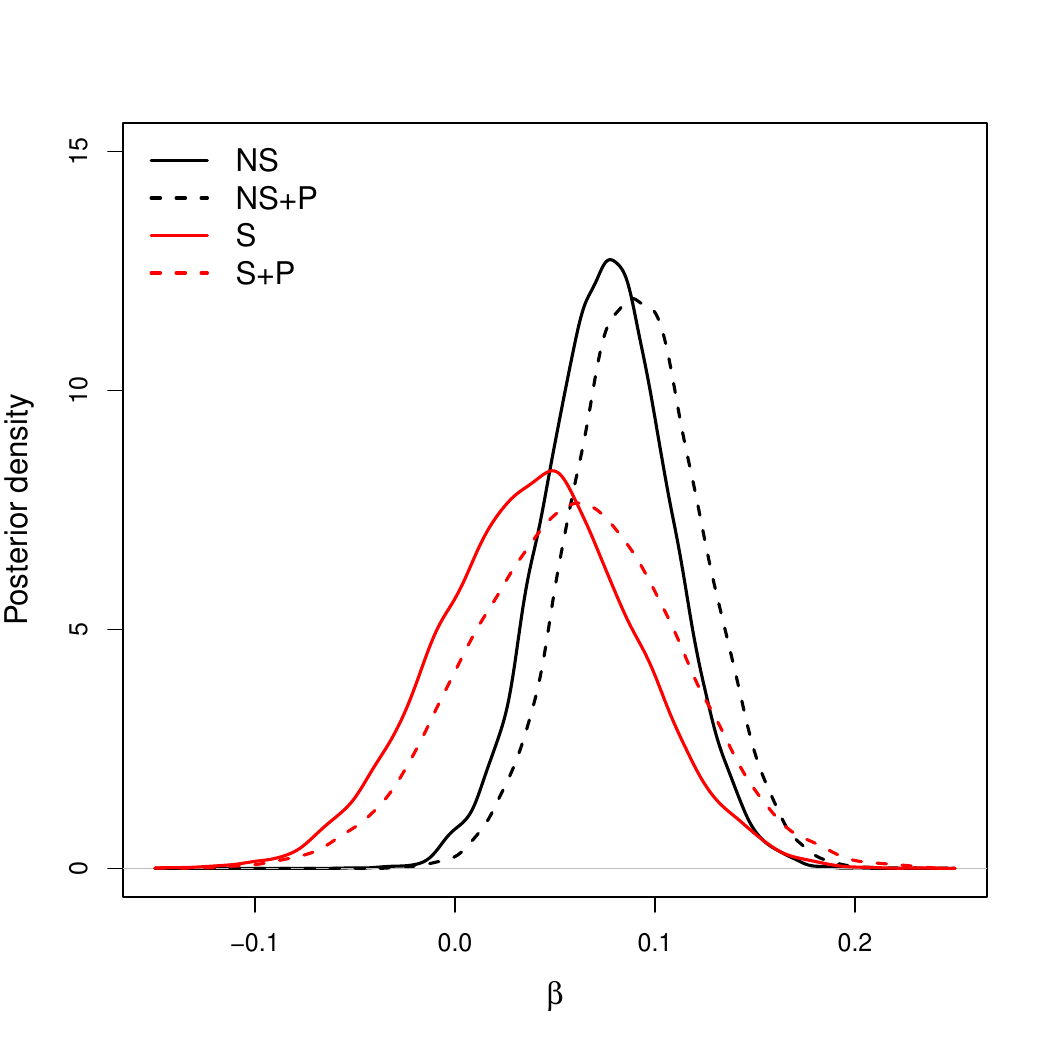}
\end{figure}

The posterior distributions of $\beta$ under these four models are plotted in Figure \ref{f:covid_beta}.   The spatial models give smaller posterior mean and larger posterior variance than the non-spatial models.  Including the generalized propensity score leads to a slightly higher effect estimate for both the spatial and non-spatial analyses.  The results are generally similar to those in \cite{wu2020exposure} who found an 8\% increase in COVID-19 related mortality for a unit increase in long-term average PM$_{2.5}$. Therefore, this analysis does not detect a missing spatial confounder that dramatically affects the causal effect estimate.

\section{Methods for spatial interference/spillover}\label{s:spillover}

Interference (also called spillover) occurs when the treatment received by one unit can affect the outcomes of other units. The ubiquitous no interference assumption in Section \ref{s:confounders:areal:PO} was first discussed in \cite{cox1958planning}, where it was referred to as ``no interaction between units'' \citep{hernan2019causal}.  In the subsequent literature, it is often simply referenced as part of SUTVA. Despite a variety of data and treatments exhibiting interference, methods that account for interference have only recently begun to proliferate in the statistics literature, in part because interference significantly complicates the potential outcomes approach and requires additional assumptions about the form of the interference. 

In this section we review the challenges associated with accounting for interference, and the current literature on this topic.  In Section \ref{s:spillover:PO} we give a general formulation of potential outcomes in the presence of interference, and define several quantities of interest under this framework.  The remainder of the section discusses different assumptions about the nature of interference and subsequent estimation methods.

\subsection{Potential outcomes framework}\label{s:spillover:PO}

In the potential outcomes framework in Section \ref{s:confounders:areal:PO} with binary treatment and no interference, there are two potential outcomes defined for each unit: $Y_{ij}(0)$ and $Y_{ij}(1)$.   Allowing for general treatment interference entails considering $2^n$ potential outcomes, each corresponding to a different combination of treatments received by all units.  As a result, the estimands under interference are more complicated because they require considering treatment that could be applied to multiple units. Therefore, defining the potential outcomes and estimands requires additional notation.  We distinguish between the treatment applied to unit $(i,j)$ in the observed dataset, $A_{ij}$, and a hypothetical treatment that could be applied to unit $(i,j)$, denoted $a_{ij}$.  To describe potential outcomes under interference we denote the treatments that could be applied to all $n$ units as $\ba=\{a_{ij}; i=1,...,N; j=1,...,n_i\}$, and the collection of the $n-1$ treatments excluding $a_{ij}$ as $\ba_{-ij}$. The potential outcome for each unit is then written as $Y_{ij}(a_{ij},\ba_{-ij})$, where the first term is the treatment received by unit $(i,j)$ and the second term are the treatments received by other units. 

The average treatment effect in (\ref{e:ATE}) is insufficient in the presence of interference as it depends only on the treatment assigned to unit $(i,j)$. Rather, several treatment effects are needed to provide a comprehensive summary.  \cite{halloran1991study, halloran1995causal} and \cite{hudgens2008toward} describe four key estimands assuming binary treatments. The direct effect (DE) is
\begin{equation}\label{e:DE}
DE_{ij}(\ba_{-ij}) = \mbox{E}\left\{Y_{ij}(1, \ba_{-ij}) - Y_{ij}(0,  \ba_{-ij})\right\}.
\end{equation}
The direct effect compares the difference potential outcomes for unit $(i,j)$ with treatments $A_{ij}=1$ versus $A_{ij}=0$ and holding all other treatments fixed at $\ba_{-ij}$.  Unlike (\ref{e:ATE}), there is not a single direct effect, as (\ref{e:DE}) may be different for each unit and for all $2^{n-1}$ combinations of $\ba_{-ij}$. 
While the direct effect isolates the local treatment effect, the indirect effect (IE) measures the contribution of other treatments,
\begin{equation}\label{e:IE}
 IE_{ij}(\ba_{-ij}, \ba_{-ij}') = \mbox{E}\left\{Y_{ij}(0, \ba_{-ij}) -
    Y_{ij}(0, \ba_{-ij}')\right\}.
\end{equation}
The indirect effect is also called the spillover effect because it compares the difference between potential outcomes for two combinations of treatments for the other units, $\ba_{-ij}$ and $\ba_{-ij}'$, to an untreated unit with $a_{ij}=0$ to quantify how much of the other treatment effects spill over to observation $(i,j)$.  The direct and indirect effects can be combined using either the total (TE) or overall effects(OE):
\begin{eqnarray}\label{e:TotEffect}
     TE_{ij}(\ba_{-ij}, \ba_{-ij}') 
     &=& DE_{ij}(\ba_{-ij}) + IE_{ij}(\ba_{-ij}, \ba_{-ij}')
     = \mbox{E}\left\{Y_{ij}(1, \ba_{-ij}) - Y_{ij}(0,\ba_{-ij}')\right\}\nonumber\\
     OE_{ij}(\ba, \ba') &=& \mbox{E}\left\{Y_{ij}(a_{ij}, \ba_{-ij}) -
    Y_{ij}(a_{ij}', \ba_{-ij}')\right\}.\nonumber
\end{eqnarray} 
These effects are similar, except that the total effect always compares $a_{ij}=1$ versus $a_{ij}=0$ whereas the overall effect allows the local treatment to be the same for $\ba$ and $\ba'$.

If these effects can be estimated, then the user can interrogate the fitted model by selecting any scenarios defined by $\ba$ and $\ba'$.  For example, in the context of Example 1, the direct effect might be computed by fixing the air pollution status of all other units $\ba_{-ij}$ at their current value to determine the effect of a local action that changes the air pollution concentration in the mother's zip code but does not affect other zip codes.  For the indirect effect we might fix all the treatment variables at their observed values except set the air pollution variable for the zip codes neighboring a mother's zip code to one in $\ba_{-ij}$ versus zero in $\ba_{-ij}'$ to determine the impact of changing the air pollution in zip codes where the mother spends some time outdoors.  The sum of these two effects is the total effect of changing the air pollution status of all zip codes in the mother's home range (her zip code and those the mother frequents).  This total effect equals the overall effect of setting $\ba={\bf 1}$ for the mother's home range, $\ba'={\bf 0}$ for the mother's home range, and both $\ba$ and $\ba'$ equal to the current value for all other zip codes.



While measures such as $DE_{ij}(\ba_{-ij})$ are useful for understanding the implications of individual actions on local outcomes, assessing the overall impact of the treatment requires averaging over units and potential actions. Rather than weight all potential actions equally, they can be assigned probabilities, $\mbox{Prob}(\ba={\tilde \ba}) = \psi({\tilde \ba})$.  The probability mass function $\psi$ is called the treatment policy.  For example, the policy-averaged expected counterfactual outcome under treatment $a_{ij}=a$ for unit $(i,j)$ is
\begin{equation}\label{e:policy_ave_Y}
  {\bar Y}_{ij}(a,\psi) = \sum_{ {\tilde \ba}_{-ij}}\mbox{E}\{Y_{ij}(a,{\tilde \ba}_{-ij})\}\mbox{Prob}(\ba_{-ij}={\tilde \ba}_{-ij}|a_{ij}=a)
\end{equation}
where the sum is over all $2^{n-1}$ possible values of $ \ba_{-ij}$ and $\mbox{Prob}(\ba_{-ij}|a_{ij}=a)$ is determined by the policy, $\psi$.  The policy-averaged direct effect for unit $(i,j)$ is then ${\bar Y}_{ij}(1,\psi)-{\bar Y}_{ij}(0,\psi)$, and the spatial average direct effect is
\begin{equation}\label{e:interference:policy}
    DE(\psi) = \frac{1}{n}\sum_{i=1}^N\sum_{j=1}^{n_i}
    {\bar Y}_{ij}(1,\psi)-{\bar Y}_{ij}(0,\psi).
\end{equation}
Policy-averaged indirect, total and overall effects have similar forms.

In the context of the environmental epidemiology study described in Example 1, a simple policy is to assume that the $a_{ij}$ are independent over units with $\mbox{Prob}(a_{ij}=1) = p$ and compute (\ref{e:interference:policy}) for several values of $p$ to understand the direct effect. A policy more tailored to anticipating short-term effects of interventions in a given region is to assume that the $a_{ij}$ are independent over units with $\mbox{Prob}(a_{ij}=1) = p_a$ if the current value of the treatment in unit $(i,j)$ is $A_{ij}=a$.  Under this policy, a zip code currently below the threshold is converted to exceed the threshold with probability $p_0$, and a zip code currently above the threshold is converted to below the threshold with probability $1-p_1$.  The policy-averaged direct, indirect and total effects can be approximated via Monte Carlo simulation for a range of $p_0$ and $p_1$ to evaluate the overall effects  of a campaign to reduce air pollution.

\begin{figure}
	\centering
	\caption{{\bf Variable dependencies under different forms of interference.} 
	Spatial location is indicated horizontally. Indirect effects are shown as dashed lines, and confounding relationships are shown as solid lines at a location and dotted lines across locations. $A$ is the treatment, $Y$ is the outcome, and $X$ is the observed confounder.}
	\label{fig:intDAGs}		
	\includegraphics[width=.9\textwidth]{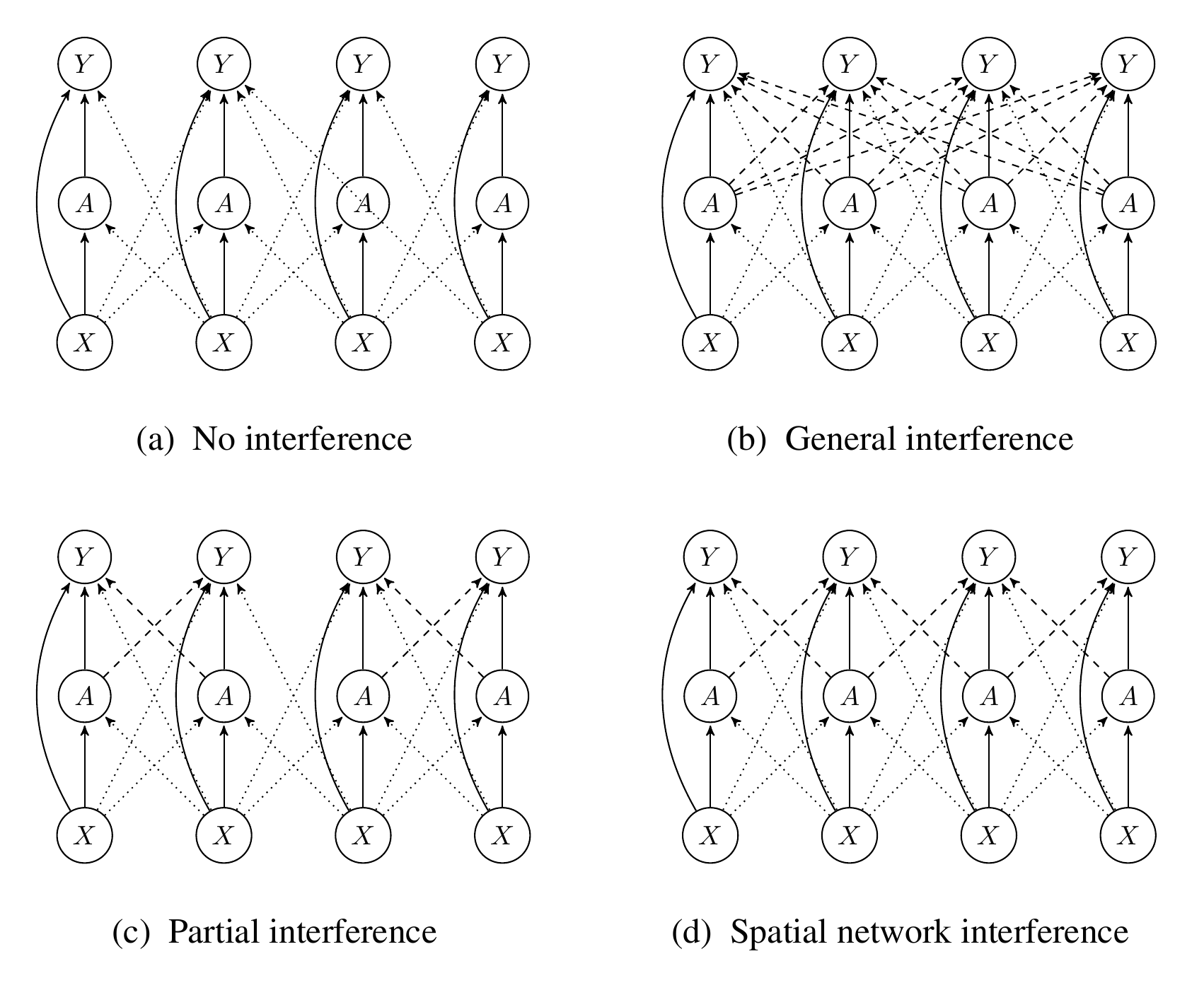}
\end{figure}

While these summaries are well defined for any potential outcome model, estimation is virtually impossible without simplifying assumptions.  In the remainder of this section we discuss several methods that exploit the spatial structure of the units to simplify the interference pattern.  These methods are summarized in Figure \ref{fig:intDAGs}.

\subsection{Partial interference}\label{s:spillover:partial}

Partial interference, a term coined in \cite{sobel2006randomized}, or clustered interference, assumes that the units can be partitioned into groups so that interference can occur only between observations in the same group. In Example 1, partial interference might be evoked if it is reasonable to partition the zip codes into cities, and that birth weight is dependent only on the air pollution concentration in the mother's city, and not air pollution in other cities.  A further parametric assumption might be that the potential outcome is a function only of the air pollution concentration in the mother's zip code and the proportion of her city's zip codes that exceed the threshold excluding zipcode $i$, denoted by ${\tilde a}_i$.  A linear model with these assumptions is 
\begin{equation}
Y_{ij}(a_{ij},\ba_{-ij}) =  a_{ij}\beta_1 + {\tilde a}_{ij}\beta_2 + \bX_{ij}\bgamma  + \varepsilon_{ij},\end{equation}
where $\beta_1$ and $\beta_2$ entail the direct and indirect effects, respectively.  This parametric model and assumptions analogous to Assumptions \ref{asump:SUTVA}, \ref{asump:LatentIgnorability} and \ref{asump:positivity} that $\bA$ is independent of all potential outcomes given the $n$ vectors $\bX_{ij}$ and that $\phi(\ba)>0$ for all $\ba$ endows the parametric model 
\begin{equation}\label{e:Y_PI}
 Y_{ij} =  A_{ij}\beta_1 + {\tilde A}_{ij}\beta_2 + \bX_{ij}\bgamma  + \varepsilon_{ij}
\end{equation}
with a causal interpretation.  Of course, this model relies on strong assumptions that are difficult to verify, and thus a more flexible approach may be desirable. 

There is an extensive literature that explores and expands on non-spatial partial interference \citep{halloran1991study, halloran1995causal,halloran2012minicommunity,tchetgen2012causal,vanderweele2014interference,liu2016inverse,barkley2017causal, baird2018optimal,papadogeorgou2019causal}.  \cite{zigler2012estimating} assume partial interference in a spatial analysis of the health effects of environmental regulations, with clusters of sites defined by their attainment status. \cite{perez2014assessing} and \cite{zigler2018bipartite} assume partial interference for groups defined by spatial proximity.  \cite{zigler2018bipartite} deal with additional complications that arise when the spatial resolutions of the treatment and response differ. 

\subsection{Spatial network interference}\label{s:spillover:network}

With the rise of social network data, there is a fast-growing literature on network-based interference, where observations can interfere with each other along connected edges.  These methods can be applied to areal spatial data by viewing the regions as the network's nodes and defining spatial adjacency by the network's edges \citep[e.g.,][]{verbitsky2012causal}.  For example, as in the CAR model defined in Section \ref{s:confound:review}, regions $i$ and $k$ can be defined as sharing an edge if they share a common border. A simple example of a model to study spatial network interference for Example 1 is (\ref{e:Y_PI}) with ${\tilde A}_{ij}$ redefined as the mean treatment variable across the $m_i$ neighbors of region $i$.

More generally, \cite{forastiere2016identification} propose a model that allows for interference
between an observation and its immediate neighbors, creating a local interference neighborhood around each observation. Treatment effects are estimated by conditioning on propensity scores for the direct and indirect treatment effects.  \cite{aronow2017estimating} considers network data in a similar vein, but loosens the restrictions on interference by defining an exposure mapping function.   \cite{tchetgen2017auto} examine arbitrary network interference subject only to a local Markov property that observations are conditionally independent after taking into account the nodes between them. This gives both a reasonable constraint for estimation and also allows for treatment effects to propagate through the network.  In a further generalization of the spatial network interference assumption, \cite{giffen2020generalized} use the distance between units themselves, rather than a network approximation, to develop a generalized propensity score method to balance the spillover effect, ${\bar A}_i$.

\subsection{Process-based spillover models}\label{s:spillover:numerical}

Partial and network interference make assumptions that are conducive to a statistical analysis, such as the simple spillover effect in (\ref{e:Y_PI}), but are likely crude representations of reality.  Mechanistic methods that encode scientific understanding of the physical processes of interest offer increased fidelity to the true interference structure. Mechanistic models are indispensable in environmental attribution studies.  For example, climate models play a central role in the Intergovernmental Panel on Climate Change's conclusion that human activities likely caused the majority of the observed increase in global mean surface temperature from 1951 to 2010 \citep{bindoff2013detection}.  As reviewed by \cite{hegerl2011use}, unlike purely statistical models that are limited to scenarios observed in the data, mechanistic models can be run under counterfactual scenarios that have not, or could not, be observed.  This provides a key link to the potential outcomes framework in Section \ref{s:spillover:PO}.

While mechanistic models can be used to estimate direct effects, they are more critical in the presence of interference because they can rule out many of the massive number of potential spillover paths, greatly reducing the complexity of the problem. Despite these strengths, mechanistic models are only approximations, and thus need to be calibrated and validated using observed data.  Most relevant for our purposes is the recent work that combines mechanistic modelling with spatial statistical methods to estimate causal effects.  For example,  \cite{Larsen2020spatial} fit a Bayesian geostatistical model to observed air pollution concentrations and mechanistic model output under scenarios with and without wildland fires to map the total causal effect of wildland fires on fine particulate matter concentration and the resulting health burden.  Rather than post-processing model runs,  \cite{forastiere2020bipartite} build a statistical model based on a dispersion model to track air pollution from power plants in a causal analysis of health effects, and \cite{cross2019confronting} embed an epidemiological model for disease spread in a hierarchical Bayesian model to estimate spillover effects.  These examples that highlight the important roles of mechanistic models not only likely provide more accurate estimates of causal effects, but also ensure the results are tethered to scientific theory.

\section{Spatiotemporal methods}\label{s:ST}

Data collected over space and time are more informative about causal relationships than cross-sectional data, because they afford the opportunity to observe variables coevolve. This reduces the potential for spurious associations.  For example, if a treatment is applied in the course of the study, comparing a site's responses before and after the treatment can control for missing spatial confounding variables assuming they and their effects are time-invariant.  This narrows the search for potential confounding variables to those with a similar pattern as the treatments over both space and time. 

To describe spatiotemporal methods, we adopt new notation to accommodate the temporal dimension.  For simplicity, we assume areal spatial units, discrete time steps, and that each region $i\in\{1,...,N\}$ has a single observation at each time step $t\in\{1,...,T\}$.  We denote the response, treatment, known and unknown confounding variables as $Y_{it}$, $A_{it}$, $\bX_{it}$ and $U_{it}$, respectively.  The potential outcomes framework and assumptions in Section \ref{s:confounders:areal:PO} apply with the time step $t$ replacing the replication number $j$.  Similarly, many of the spatial methods in Section \ref{s:confound} such as matching (Section \ref{s:confounders:areal:matching}), neighborhood adjustments (Section \ref{s:confounders:areal:neighborhood}), propensity score methods (Section \ref{s:confounders:areal:propscores}) and the instrumental variable approach (Section \ref{s:confounders:areal:IM}) apply for spatiotemporal data by viewing time as a third spatial dimension, with a different degree of correlation in this third dimension.

\subsection{Testing for missing spatial confounders}

\cite{janes2007trends} propose a method to test for unmeasured spatial confounders using spatiotemporal data.  Letting ${\bar A}_t$ denote the average of $A_{it}$ at time $t$, their approach can be adapted to our setting via the model
\begin{equation}\label{e:janes}
Y_{it} = \eta_1{\bar A}_{t}+ \eta_2(A_{it}-{\bar A}_{t}) + \bX_{it}\bgamma + \varepsilon_{it} 
\end{equation}
where $\bX_{it}$ includes smooth functions of $t$ to account for missing temporally-varying confounders.  In this model, $\eta_1$ and $\eta_2$ measure global and local effects of treatment, respectively, and they argue that if the estimated values of $\eta_1$ and $\eta_2$ are equal and non-zero then this represents an average causal effect of $A_{it}$ on $Y_{it}$, and that a large difference between the estimated $\eta_1$ and $\eta_2$ suggests there may be a missing spatial confounder.

\subsection{Difference in difference methods}\label{s:ST:DID}  
Difference-in-difference (DID) estimators \citep{ashenfelter1984using} aim to quantify the treatment effect on the increase in the mean response over time.  For simplicity we assume a binary treatment variable and two time steps ($T=2$).   If the treatment at the both time steps is $a_{i1}=a_{i2}=a$, the increase in counterfactuals at site $i$ is $\delta_i(a) = Y_{i2}(a)-Y_{i1}(a)$.  Therefore, $\delta_i(0)$ is the increase over time in the absence of treatment, and $\delta_i(1)-\delta_i(0)$ is the increase that can be attributed to treatment.  The DID average treatment effect is then
\begin{equation}\label{e:ST:DID1}
\delta^{DID} = \mbox{E}\left[\frac{1}{N}\sum_{i=1}^N\{\delta_i(1)-\delta_i(0)\}\right],
\end{equation}
which is analogous to (\ref{e:ATE}) except 
that the outcomes are changes over time. Assume the potential outcomes follow the model $Y_{it}(a) = \beta_1a + \beta_2t + \beta_3ta + \bX_{it}\bgamma + U_{it} + \varepsilon_{it}$.  Under Assumptions \ref{asump:SUTVA}--\ref{asump:positivity}, the observed outcome model follows the induced linear model
\begin{equation}
\label{eq:did}
Y_{it} = \beta_1A_{it} + \beta_2t + \beta_3tA_{it} + \bX_{it}\bgamma + U_{it} + \varepsilon_{it}.
\end{equation}
Moreover, $\beta_3=\delta^{DID}$ has a causal interpretation. 

To render Assumptions 1--3 plausible, it is important to include information on a rich enough set of time-varying confounders in $\bX_{it}$ that affect both $A_{it}$ and $Y_{it}$. In the spatiotemporal settings, the time-varying confounders $\bX_{it}$ include the observed information on the past treatments and outcomes.

\cite{delgado2015difference} extend the spatial DIDs  by assuming Markov interference where treatment effects only impact neighbors. This gives the model
\begin{equation}
\label{eq:did:spillover}
Y_{it} = \beta_1A_{it} + \beta_2t + \beta_3tA_{it} + \beta_4{\bar A}_{it} + \beta_5t{\bar A}_{it} + \bX_{it}\bgamma + U_{it} + \varepsilon_{it}
\end{equation}
where  ${\bar A}_{it}$ is the mean of $A_{it}$ over the $m_i$ neighbors of region $i$ at time step $t$. The neighborhood coefficients $\beta_4$ and $\beta_5$ can be viewed either as indirect spillover effects or added terms to adjust for local confounders to give more precise estimates of the direct causal effect, $\beta_3$. 

Matched wake analysis combines the DID approach with a spatiotemporal analogue to coarsened exact matching \citep{schutte2014matched}. It was developed in the political science literature for studying responses to whether insurgent violence in Iraq causes civilians to help the US military. In this scenario, insurgent violence leading to civilian casualties is the ``treatment`` and violence not resulting in casualties is the ``control''. The response is the act of turning in salvaged unexploded ordinance to the US military, so that it will not be used in an improvised explosive device. The data are divided into sliding spatiotemporal windows called ``wakes" and matched. Then a difference-in-differences approach is applied to the matched sample by counting the number of explosives turned in before and after events. A drawback to this method is that in some cases the sliding windows may overlap, which will violate SUTVA. 

\subsection{Granger causality}\label{s:ST:Granger}

Granger causality is a fundamentally different concept than the potential outcomes framework.  It is defined by temporal relationships and not potential outcomes.  In a time series analysis with response $Y_t$, treatment $A_t$, and all other relevant variables at time $t$, $\bX_t$, the treatment is said to Granger cause the response if $\mbox{Var}(Y_t|\bH_t)>\mbox{Var}(Y_t|\bH_t,A_1,...,A_{t-1})$, where the history up to time $t$ is $\bH_t=\{Y_1,...,Y_{t-1},\bX_1,...,\bX_{t-1}\}$.  In other words, Granger causality implies that given the history of all other variables, knowledge of past treatments reduces predictive uncertainty.  If a linear lag $L$ time series model is assumed, $Y_{t} = \sum_{l=l}^L(A_{t-l}\beta_l + \bX_{t-l}\bgamma_l + Y_{t-l}\rho_l) + \varepsilon_t$, then the treatment is said to Granger cause the response if $\beta_l\ne 0$ for any $l\in\{1,...,L\}$. 

Because this notion of causality is inherently defined for temporal data, extending these methods to the spatiotemporal case is straightforward.  The simplest model is the linear no-interference model
\begin{equation}\label{e:ST:granger1}
Y_{it} = \sum_{l=l}^L\left\{A_{it-l}\beta_l + \bX_{it-l}\bgamma_l + Y_{it-l}\rho_l\right\}+U_{it}+\varepsilon_{it},
\end{equation}
where $U_{it}$ is correlated over space (e.g., following a CAR or SAR distribution) but independent over time.  It is also straightforward to include spillover effects by including spatial averages as covariates, i.e., under a Markov interference assumption the mean of $A_{it-1}$ over region $i$'s $m_i$ neighbors could be added as a covariate.

Granger causality and Rubin causality based on potential outcomes are fundamentally different. Granger causality is defined in terms of predictive uncertainty, as might be useful to a passive observer of the system trying to maximize predictive power.  In contrast, Rubin causality is defined in terms of the effects of an active intervention, as might be performed by a scientist conducting a controlled experiment.  Despite their different definitions and objectives, these two approaches share similarities.  \cite{white2010granger} show  that Granger causality is equivalent to Rubin causality for times series data with no missing confounders and valid parametric assumptions. For example, the model in (\ref{e:ST:granger1}) could be motivated by Granger causality or Rubin causality with Assumptions 1-4 and further assumptions (normality, linearity, etc) on the form of the potential outcomes model.  For further discussion of the similarities and differences between types of causality, see \cite{holland1986statistics} or \cite{eichler2012causal}.

\section{Methods for point-referenced data}\label{s:geostat}

Point-referenced, or geostatistical, data are not measurements of a region, but rather taken at a specific point (latitude/longitude).   Let $\bs_{i}\in\calR^2$ be the spatial location corresponding to observation $i\in\{1,...,n\}$.  The spatial regression model becomes
\begin{equation}\label{e:geostat}
 Y_i = A_i\beta + \bX_i\bgamma + U(\bs_i) + \varepsilon_i
\end{equation}
where the unknown confounder $U(\bs)$ is a spatial processes and $\varepsilon_i\iid\mbox{Normal}(0,\tau^2)$.  This notation allows for replications at sites if, say $\bs_i=\bs_j$, in which case observations $i$ and $j$ share the spatial term $U(\bs_i)=U(\bs_j)$. The covariate vector $\bX_i$ can include spatial covariates such as the elevation at $\bs_i$ and  non-spatial covariates such as the time of day the measurement was taken. 

Unlike an areal data analysis as in Section \ref{s:confound} where the number of potential sampling locations is finite, a geostatistial analysis must consider an uncountable number of potential sampling locations $\bs\in\calD\subset \calR^2$.  We use the bold to denote a process over the entire spatial domain; e.g., $\bU=\{U(\bs):\bs\in\calD \}$.  An unknown spatial process such as $\bU$ is typically assumed to be a continuous function of $\bs$ over $\calD$ and modeled as a Gaussian process with mean zero and isotropic covariance function (i.e., a covariance that depends only on the distance between locations).  Although other covariance functions are available \citep{banerjee2014hierarchical}, the simplest choice is the exponential covariance function $\mbox{Cov}\{U(\bs_i),U(\bs_j)\} = \sigma^2\exp(-d_{ij}/\rho)$ where $d_{ij}$ is the distance between $\bs_i$ and $\bs_j$.  We denote this Gaussian process model as $U\sim\mbox{GP}(\rho,\sigma)$.  

\subsection{Potential outcomes framework}\label{s:Geo:potential}

In the most general form, the potential outcomes for observation $i$ depend on the entire spatial field of potential treatments, $\ba = \{a(\bs):\bs\in\calD\}$. Therefore, we define the potential outcome for observation $i$ as $Y_i(\ba)$.  In the context of Example 1, $a(\bs)$ might be the air pollution concentration at spatial location $\bs$, as opposed to the average concentration in a zip code.  In this geostatistical setting, a mother's exposure to air pollution would integrate the concentration $a(\bs)$ along the path the mother travels.  This could be estimated by a backpack the mother wears that continuously measures her local air pollution concentration.  Therefore, changing $a(\bs)$ for any $\bs$ in the spatial domain could affect her potential outcome.  

The potential outcomes framework simplifies dramatically under the no interference assumption.  With a binary treatment, the two potential outcomes for unit $i$ are $Y_i(0)$ if $a(\bs_i)=0$ and $Y_i(1)$ if $a(\bs_i)=1$.  In this simple case, the potential outcomes concepts, definitions and assumptions introduced in Section \ref{s:confounders:areal:PO} directly apply to the geostatistical setting.  Many of the methods developed to adjust for missing spatial confounders described for areal data can also be applied.  For example, all of the propensity score methods in Section \ref{s:confounders:areal:propscores} and instrumental variables methods in Section \ref{s:confounders:areal:IM} can be adapted for geostatistical data by replacing the CAR model for the missing spatial confounder with a Gaussian process model.  Many of the other methods introduced for areal data can also be modified for geostatistical applications, as described in the remainder of this section.

\subsection{Matching methods}\label{s:Geo:matching}
The matching methods described in Section \ref{s:confounders:areal:matching} that pair observations from the same region can be applied for geostatistical data with replications at spatial locations. Distance adjusted propensity score matching (DAPSm) \citep{papadogeorgou2018adjusting} can be used when there are not replications. This method alters propensity score matching \citep{rosenbaum&rubin83a} by using a standardized distance that combines the propensity score difference and geographic distance.  The logic is that if unmeasured spatial confounders exist, then observations that are close together will have confounders that are the most alike. Similar to the neighborhood adjustment methods, this method balances treatment and control by including geographic distances as a proxy for the unmeasured confounders in the matching process. The difference for a pair with $A_i=1$ and $A_j=0$ is defined as
\begin{equation}
 D_{ij} = w|{\hat e}_i - {\hat e}_j| + (1-w)d_{ij}/m
\end{equation}
where ${\hat e}_i$ and ${\hat e}_j$ are estimated propensity scores, $m$ is the maximum distance between pairs of locations in the study domain and $w \in [0,1]$ is a weight. The authors propose an algorithm to select pairs with small $D_{ij}$. 

\subsection{Regression discontinuity}

\begin{figure}
	\centering
	\caption{{\bf Illustration of regression discontinuity.} The treatment region $\calA$ is the region above the curve, the points are the sample locations $\bs_i$ with samples with $A_i=1$ filled and the background color is the mean function $A(\bs)\beta+U(\bs)$ were $A(\bs)$ indicates that $\bs=(s_1,s_2)\in\calA$.} 
	\label{fig:RegDist}		\includegraphics[width=.5\textwidth]{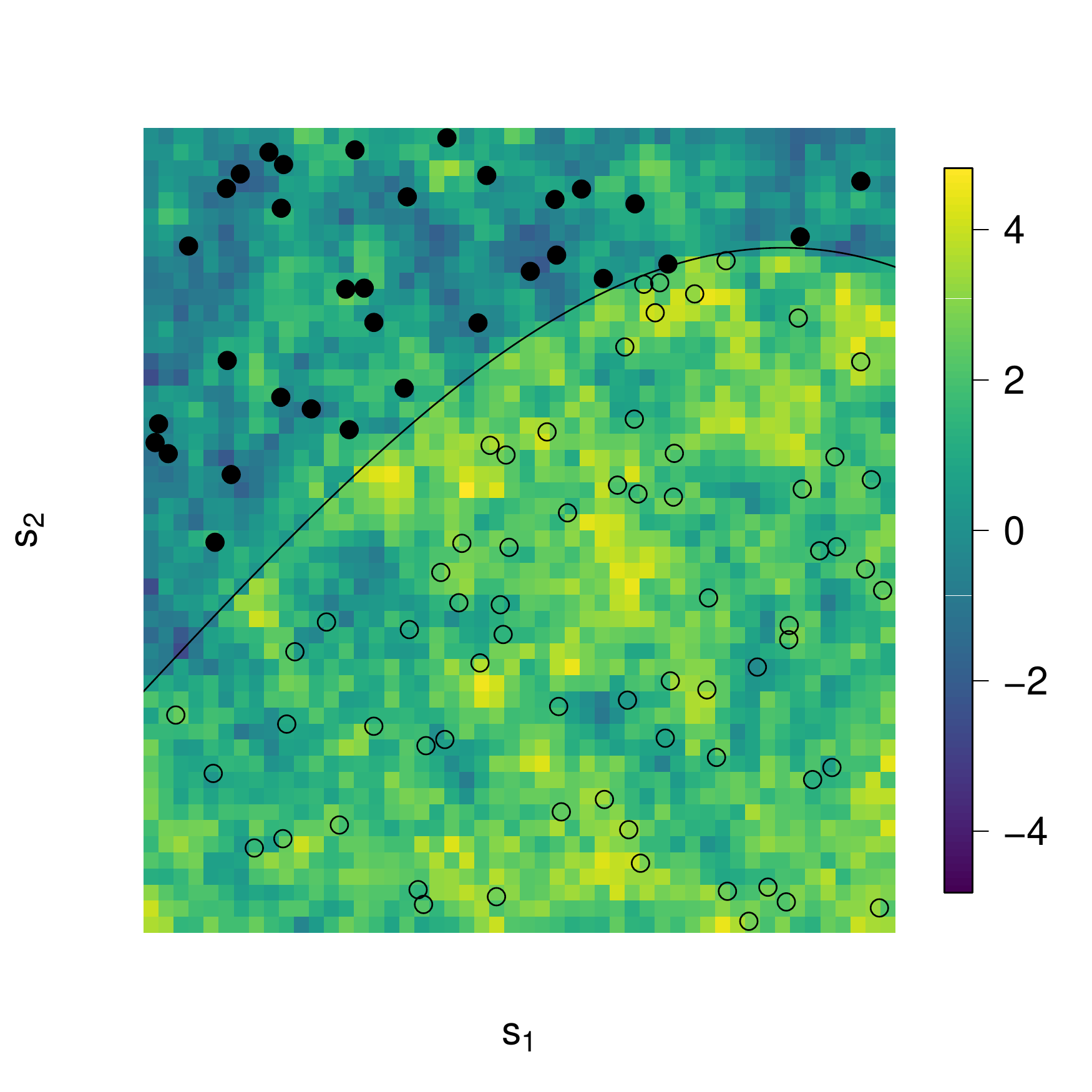}
\end{figure}

Regression discontinuity designs are generally used when treatment assignment is determined by whether the covariate value for a unit exceeds a threshold
\citep{imbens2008regression,bor2014regression,keele2015geographic}, e.g., students are admitted to a college if and only if their SAT score exceeds a threshold.  These cases provide a natural experiment if it can be assumed that units slightly above and slightly below the threshold are similar in every way except the treatment assignment, and thus the difference between these groups can be attributed to the causal effect of the treatment.  Natural experiments of this form often arise in environmental and epidemiological studies where the variable being thresholded to determine treatment is the spatial location.  In the context of Example 1, the treatment might be whether a state is subject to an air pollution regulation, and the objective is to determine if this affects health outcomes.  Figure \ref{fig:RegDist} shows a hypothetical example where treatment is applied to locations in the region $\bs\in\calA\subset\calD$.  If it can be assumed that all other factors are balanced across the border of $\calA$, then comparing observations on either side of the border provides information about the causal effect of treatment. Under this assumption, the causal effect can be estimated by simply fitting the geostatistical model in (\ref{e:geostat}) with $A_i = 1$ if $\bs_i\in\calA$ and $A_i=0$ otherwise.  

\subsection{Neighborhood adjustments}\label{s:Geo:neighborhood}

\subsubsection{Stochastic partial differential equation modeling}

Section \ref{s:confounders:areal:neighborhood} introduces the SAR model that defines the regression of the response onto the treatment after subtracting the means across neighboring regions. The motivation for building a model on the differences is to remove the effects of spatially-smooth confounding variables.  The stochastic partial differential equation (SPDE) models of \cite{lindgren2011explicit} can be viewed as an extension of this idea to the continuous (geostatistical) spatial domain.   In the SPDE framework, models are specified on the partial derivatives of the response surface, which is a generalization of the SAR model that can be applied to differentiable  functions such as $\bU$.  \cite{lindgren2011explicit} show that this approach can be used to approximate Gaussian processes with the $\Matern$ covariance function, and develop approximations that resemble the SAR covariance model.  



\subsection{Spillover/interference methods}\label{s:Geo:splillover}

Defining interference for geostatistical applications requires returning to the general potential outcomes formulation in Section \ref{s:Geo:potential}, where the potential outcome for observation $i$ depends on the entire field of treatments, $\ba$, and is denoted as $Y_i(\ba)$.
Relating the spatial field $\ba$ with the scalar potential outcome requires assumptions about the form of interference.  A general form of the interference is
\begin{eqnarray}
  Y_i(\ba) &=& a(\bs_i)\beta_1 + {\bar a}_i\beta_2+\bX_i\bgamma+U_i + \varepsilon_i\label{e:geostat:intY}\\
  {\bar a}_i &=& \int_{\calD}w(\bs_i,\bs)a(\bs)d\bs\label{e:geostat:intA},
\end{eqnarray}
where $w$ is a weighting function that determines the spillover effect ${\bar a}_i$ and $\beta_1$ and $\beta_2$ control the direct and indirect effects, respectively.  Given this potential outcome model, the four causal effects (direct, indirect, total and overall) can be defined and interpreted as in Section \ref{s:spillover:PO} with $\ba_{-i}$ defined as the surface $\ba$ excluding $a(\bs_i)$, or perhaps excluding $\ba$ for all sites within a small radius of $\bs_i$.

The form of spillover in (\ref{e:geostat:intA}) encompasses many common interference assumptions.  For example, partial/cluster interference can be implemented by fixing $w(\bs_i,\bs)=0$ if sites $\bs_i$ and $\bs$ are in different groups.  A structure resembling Markov/network interference assumes that $w(\bs_i,\bs)=1/(\pi r^2)$ if $\bs$ is within radius $r$ of $\bs_i$ and $w(\bs_i,\bs)=0$ otherwise.  This reduces the spillover measure ${\bar a}_i$ to the average treatment within radius $r$ of $\bs_i$. If strict bounds on the range of interference cannot be assumed, then the weight function could be a decreasing function of the distance from $\bs_i$, such as the Gaussian kernel function with $w(\bs_i,\bs) = \exp\left\{-0.5(||\bs-\bs_i||/\phi)^2\right\}/\sqrt{2\pi\phi^2}$.  

Even after reducing the complexity of the model by selecting a simple form for the weighting function, computing the spatial integral in (\ref{e:geostat:intA}) is often impossible because the treatments are only observed at a finite number of locations.  One remedy is to use spatial interpolation (Kriging) to impute the treatments onto a fine grid of locations covering the spatial domain and then approximate the integrals as sums over the grid points.  In this case, uncertainty about the estimated spillover variables should be accounted for using Bayesian or multiple imputation methods.  

Given a form of interference and the assumption of no missing confounders, estimation of the direct and indirect effects can proceed with the usual spatial linear model. One approach to accounting for missing spatial confounders is to include spatial propensity score models for both the direct treatment $A_i$ and the spillover effect ${\bar A}_i$.  The propensity score for $A_i$ can be estimated as in the areal case with say a spatial logistic regression to give ${\hat e}(\bs_i)$.  

\section{Summary and future work}\label{s:discussion}

The field of spatial causal inference has seen impressive advances in recent years.  There are now methods to address the fundamental problems including accounting for missing spatial confounding variables and modeling spatial interference.  However, there are many opportunities for future work that we discuss below, including combining data types, relaxing model assumptions, going beyond mean estimation, and using causal estimates for decision making.

We have discussed methods for areal data (Section \ref{s:confound}) and point-referenced/geostatistical data (Section \ref{s:geostat}) separately, but many analyses require utilizing both types of data.  For example, treatments may be defined at point locations (e.g., air pollution concentration) while the response variable is defined regionally (e.g., hospital admission rate by zip code).  In spatial statistics this is referred to as the change of support problem \citep{gotway2002combining,gelfand2010handbook}.  One approach to combining data with different supports is to conceptualize the areal data as an aggregation of a continuous latent process and then specify geostatistical models such as those presented in Section \ref{s:geostat} on the latent process.  Extending these methods to the causal inference would require carefully specifying the causal estimand and devising computationally-efficient methods for estimation.  \cite{zigler2018bipartite} may provide a template for this work.

Change of support issues also arise when the treatment is a point source, such as an oil spill, power plant or wildland fire. The effect of point source treatment variables can be direct, but their most prominent causal effects will likely be the spillover effects (Section \ref{s:spillover}) felt by nearby locations.  The spillover effects can be modelled as a function of the distance from the response location to the point source or mechanistically using a mathematical dispersion model (Section \ref{s:spillover:numerical}).  These methods can also be extended to the spatiotemporal setting using spillover effects that decay in space and time \citep[e.g.,][]{kim2018causal,kim2019modeling}. Inferential methods that rely on modeling the treatment variables (e.g., propensity scores) could apply a spatial point pattern analysis \citep{baddeley2015spatial}, such as an inhomogeneous Poisson process model, to estimate the treatment intensity.  It may also be possible to leverage work on informative sampling \citep{diggle2010geostatistical,pati2011bayesian} that uses a joint model for the sampling locations and the responses to reduce the effects of systematic bias in the sampling design.   

Most of the methods discussed in this review rely on strong parametric assumptions such as linearity and normality.  Parametric methods dominate spatial statistics because in the canonical problem with one observation at each spatial location there is insufficient data to relax these assumptions. In contrast, most causal inference methods aim to be robust to model misspecification.  There is a body of work on nonparametric spatial methods \citep{gelfand2010handbook,reich2015spatial} that might be used to relax the parametric assumptions in spatial causal inference, but these ideas have yet to be applied in this context.

We focused only on the average treatment effect, and future work is to extend spatial causal inference to other types of treatment effects.  For example, extreme events are often the most impactful in environmental studies, and thus it would be of great interest to extend causal inference ideas to spatial quantile regression \cite[e.g.,][]{reich2011bayesian,reich2012spatiotemporal,lum2012spatial} or extreme value analysis \citep[e.g.,][]{davison2019spatial}.   Another simplification made throughout the review is that the confounder and treatment effects are the same throughout the spatial domain.  A more general approach is a locally-adaptive model with spatially-varying coefficients \citep{gelfand2003spatial}, which would be a spatial application of conditional treatment effects.

Ultimately, causal effect estimates can be used to influence decision making. An area of future work is to use these estimates to derive individualized/localized treatment rules.  This is complicated in the spatial case by interference between regions that require considering simultaneously assigning the treatments to all regions to achieve optimality. \cite{laber2018optimal} and \cite{Guan2020spatiotemporal} propose a policy-search method for optimal treatment allocation for spatiotemporal problems, but a general theory awaits development.

\section*{Acknowledgements}

This work was partially supported by the National Institutes of Health (R01ES031651-01,R01ES027892-01) and King Abdullah University of Science and Technology (3800.2).  The research described in this article has been reviewed by the Center for Public Health and Environmental Assessment, U.S. Environmental Protection Agency and approved for publication. Approval does not signify that the contents necessarily reflect the views and the policies of the Agency, nor does mention of trade names of commercial products constitute endorsement or recommendation for use. The authors declare that they have no conflict of interest.
 
\begin{singlespace}
	\bibliographystyle{rss}
	\bibliography{refs}

\begin{thebibliography}{93}
\expandafter\ifx\csname natexlab\endcsname\relax\def\natexlab#1{#1}\fi
\expandafter\ifx\csname url\endcsname\relax
  \def\url#1{\texttt{#1}}\fi
\expandafter\ifx\csname urlprefix\endcsname\relax\def\urlprefix{URL}\fi

\bibitem[{Abadie and Imbens(2006)}]{abadie2006large}
Abadie, A. and Imbens, G.~W. (2006) Large sample properties of matching
  estimators for average treatment effects.
\newblock \textit{Econometrica}, \textbf{74}, 235--267.

\bibitem[{Abadie and Imbens(2016)}]{abadie2016matching}
--- (2016) Matching on the estimated propensity score.
\newblock \textit{Econometrica}, \textbf{84}, 781--807.

\bibitem[{Aronow et~al.(2017)Aronow, Samii et~al.}]{aronow2017estimating}
Aronow, P.~M., Samii, C. et~al. (2017) Estimating average causal effects under
  general interference, with application to a social network experiment.
\newblock \textit{The Annals of Applied Statistics}, \textbf{11}, 1912--1947.

\bibitem[{Ashenfelter and Card(1985)}]{ashenfelter1984using}
Ashenfelter, O. and Card, D. (1985) Using the longitudinal structure of
  earnings to estimate the effect of training programs.
\newblock \textit{The Review of Economics and Statistics}, \textbf{67},
  648--660.

\bibitem[{Baddeley et~al.(2015)Baddeley, Rubak and
  Turner}]{baddeley2015spatial}
Baddeley, A., Rubak, E. and Turner, R. (2015) \textit{Spatial point patterns:
  methodology and applications with R}.
\newblock Chapman and Hall/CRC.

\bibitem[{Baird et~al.(2018)Baird, Bohren, McIntosh and
  {\"O}zler}]{baird2018optimal}
Baird, S., Bohren, J.~A., McIntosh, C. and {\"O}zler, B. (2018) Optimal design
  of experiments in the presence of interference.
\newblock \textit{Review of Economics and Statistics}, \textbf{100}, 844--860.

\bibitem[{Banerjee et~al.(2014)Banerjee, Carlin and
  Gelfand}]{banerjee2014hierarchical}
Banerjee, S., Carlin, B.~P. and Gelfand, A.~E. (2014) \textit{Hierarchical
  modeling and analysis for spatial data}.
\newblock Chapman and Hall/CRC.

\bibitem[{Bang and Robins(2005)}]{bang2005doubly}
Bang, H. and Robins, J.~M. (2005) Doubly robust estimation in missing data and
  causal inference models.
\newblock \textit{Biometrics}, \textbf{61}, 962--973.

\bibitem[{Barkley et~al.(2017)Barkley, Hudgens, Clemens, Ali and
  Emch}]{barkley2017causal}
Barkley, B.~G., Hudgens, M.~G., Clemens, J.~D., Ali, M. and Emch, M.~E. (2017)
  Causal inference from observational studies with clustered interference.
\newblock \textit{arXiv preprint arXiv:1711.04834}.

\bibitem[{Bind(2019)}]{bind2019causal}
Bind, M.-A. (2019) Causal modeling in environmental health.
\newblock \textit{Annual Review of Public Health}, \textbf{40}, 23--43.

\bibitem[{Bindoff et~al.(2013)Bindoff, Stott, AchutaRao, Allen, Gillett,
  Gutzler, Hansingo, Hegerl, Hu, Jain et~al.}]{bindoff2013detection}
Bindoff, N.~L., Stott, P.~A., AchutaRao, K.~M., Allen, M.~R., Gillett, N.,
  Gutzler, D., Hansingo, K., Hegerl, G., Hu, Y., Jain, S. et~al. (2013)
  \textit{{Detection and attribution of climate change: From global to
  regional}}.
\newblock Cambridge University Press.

\bibitem[{Bor et~al.(2014)Bor, Moscoe, Mutevedzi, Newell and
  B{\"a}rnighausen}]{bor2014regression}
Bor, J., Moscoe, E., Mutevedzi, P., Newell, M.-L. and B{\"a}rnighausen, T.
  (2014) Regression discontinuity designs in epidemiology: causal inference
  without randomized trials.
\newblock \textit{Epidemiology (Cambridge, Mass.)}, \textbf{25}, 729.

\bibitem[{Cao et~al.(2009)Cao, Tsiatis and Davidian}]{cao2009improving}
Cao, W., Tsiatis, A.~A. and Davidian, M. (2009) Improving efficiency and
  robustness of the doubly robust estimator for a population mean with
  incomplete data.
\newblock \textit{Biometrika}, \textbf{96}, 723--734.

\bibitem[{Cox(1958)}]{cox1958planning}
Cox, D.~R. (1958) \textit{Planning of Experiments.}
\newblock Wiley.

\bibitem[{Cross et~al.(2019)Cross, Prosser, Ramey, Hanks and
  Pepin}]{cross2019confronting}
Cross, P.~C., Prosser, D.~J., Ramey, A.~M., Hanks, E.~M. and Pepin, K.~M.
  (2019) {Confronting models with data: The challenges of estimating disease
  spillover}.
\newblock \textit{{Philosophical Transactions of the Royal Society B}},
  \textbf{374}, 20180435.

\bibitem[{Davis et~al.(2019)Davis, Neelon, Nietert, Hunt, Burgette, Lawson and
  Egede}]{davis2019addressing}
Davis, M.~L., Neelon, B., Nietert, P.~J., Hunt, K.~J., Burgette, L.~F., Lawson,
  A.~B. and Egede, L.~E. (2019) {Addressing geographic confounding through
  spatial propensity scores: A study of racial disparities in diabetes}.
\newblock \textit{Statistical Methods in Medical Research}, \textbf{28},
  734--748.

\bibitem[{Davison and Huser(2019)}]{davison2019spatial}
Davison, A.~C. and Huser, R. (2019) \textit{Spatial extremes}.
\newblock CRC Press.

\bibitem[{Delgado and Florax(2015)}]{delgado2015difference}
Delgado, M.~S. and Florax, R.~J. (2015) Difference-in-differences techniques
  for spatial data: Local autocorrelation and spatial interaction.
\newblock \textit{Economics Letters}, \textbf{137}, 123--126.

\bibitem[{Diggle et~al.(2010)Diggle, Menezes and Su}]{diggle2010geostatistical}
Diggle, P.~J., Menezes, R. and Su, T.-l. (2010) Geostatistical inference under
  preferential sampling.
\newblock \textit{Journal of the Royal Statistical Society: Series C (Applied
  Statistics)}, \textbf{59}, 191--232.

\bibitem[{Eichler(2012)}]{eichler2012causal}
Eichler, M. (2012) Causal inference in time series analysis.
\newblock In \textit{Causality: Statistical Perspectives and Applications} (ed.
  L.~B. Carlo~Berzuini, Philip~Dawid), chap.~22, 326--354. Wiley Online
  Library, 1 edn.

\bibitem[{Forastiere et~al.(2016)Forastiere, Airoldi and
  Mealli}]{forastiere2016identification}
Forastiere, L., Airoldi, E.~M. and Mealli, F. (2016) Identification and
  estimation of treatment and interference effects in observational studies on
  networks.
\newblock \textit{arXiv preprint arXiv:1609.06245}.

\bibitem[{Forastiere et~al.(2020+)Forastiere, Mealli and
  Zigler}]{forastiere2020bipartite}
Forastiere, L., Mealli, F. and Zigler, C. (2020+) {Bipartite interference and
  air pollution transport: Estimating health effects of power plant
  interventions}.
\newblock Submitted.

\bibitem[{Gelfand et~al.(2010)Gelfand, Diggle, Guttorp and
  Fuentes}]{gelfand2010handbook}
Gelfand, A.~E., Diggle, P., Guttorp, P. and Fuentes, M. (2010) \textit{Handbook
  of Spatial Statistics}.
\newblock CRC Press.

\bibitem[{Gelfand et~al.(2003)Gelfand, Kim, Sirmans and
  Banerjee}]{gelfand2003spatial}
Gelfand, A.~E., Kim, H.-J., Sirmans, C. and Banerjee, S. (2003) Spatial
  modeling with spatially varying coefficient processes.
\newblock \textit{Journal of the American Statistical Association},
  \textbf{98}, 387--396.

\bibitem[{Giffin et~al.(2020)Giffin, Reich, Yang and
  Rappold}]{giffen2020generalized}
Giffin, A., Reich, B.~J., Yang, S. and Rappold, A.~G. (2020) Generalized
  propensity score approach to causal inference with spatial interference.
\newblock \textit{arXiv preprint arXiv:2007.00106}.

\bibitem[{Gotway and Young(2002)}]{gotway2002combining}
Gotway, C.~A. and Young, L.~J. (2002) Combining incompatible spatial data.
\newblock \textit{Journal of the American Statistical Association},
  \textbf{97}, 632--648.

\bibitem[{Granger(1969)}]{granger1969investigating}
Granger, C.~W. (1969) Investigating causal relations by econometric models and
  cross-spectral methods.
\newblock \textit{Econometrica: Journal of the Econometric Society}, 424--438.

\bibitem[{Guan et~al.(2020)Guan, Reich and Laber}]{Guan2020spatiotemporal}
Guan, Q., Reich, B.~J. and Laber, E.~B. (2020) {A spatiotemporal recommendation
  engine for malaria control}.
\newblock \textit{arXiv preprint arXiv:2003.05084}.

\bibitem[{Halloran(2012)}]{halloran2012minicommunity}
Halloran, M.~E. (2012) The minicommunity design to assess indirect effects of
  vaccination.
\newblock \textit{Epidemiologic methods}, \textbf{1}, 83--105.

\bibitem[{Halloran and Struchiner(1991)}]{halloran1991study}
Halloran, M.~E. and Struchiner, C.~J. (1991) Study designs for dependent
  happenings.
\newblock \textit{Epidemiology}, \textbf{2}, 331--338.

\bibitem[{Halloran and Struchiner(1995)}]{halloran1995causal}
--- (1995) Causal inference in infectious diseases.
\newblock \textit{Epidemiology}, \textbf{6}, 142--151.

\bibitem[{Hanks et~al.(2015)Hanks, Schliep, Hooten and
  Hoeting}]{hanks2015restricted}
Hanks, E.~M., Schliep, E.~M., Hooten, M.~B. and Hoeting, J.~A. (2015)
  Restricted spatial regression in practice: geostatistical models,
  confounding, and robustness under model misspecification.
\newblock \textit{Environmetrics}, \textbf{26}, 243--254.

\bibitem[{Hansen(2004)}]{hansen2004full}
Hansen, B.~B. (2004) Full matching in an observational study of coaching for
  the sat.
\newblock \textit{Journal of the American Statistical Association},
  \textbf{99}, 609--618.

\bibitem[{He(2018)}]{he2018inverse}
He, Z. (2018) Inverse conditional probability weighting with clustered data in
  causal inference.
\newblock \textit{arXiv preprint arXiv:1808.01647}.

\bibitem[{Heckman et~al.(1997)Heckman, Ichimura and Todd}]{heckman1997matching}
Heckman, J.~J., Ichimura, H. and Todd, P.~E. (1997) Matching as an econometric
  evaluation estimator: Evidence from evaluating a job training programme.
\newblock \textit{Rev. Econ. Stud.}, \textbf{64}, 605--654.

\bibitem[{Hegerl and Zwiers(2011)}]{hegerl2011use}
Hegerl, G. and Zwiers, F. (2011) Use of models in detection and attribution of
  climate change.
\newblock \textit{Wiley interdisciplinary reviews: climate change}, \textbf{2},
  570--591.

\bibitem[{Hernán and Robins(2020)}]{hernan2019causal}
Hernán, M.~A. and Robins, J.~M. (2020) \textit{{Causal inference: What if}}.
\newblock Boca Raton: Chapman \& Hall/CRC.

\bibitem[{Hirano and Imbens(2004)}]{hirano2004propensity}
Hirano, K. and Imbens, G.~W. (2004) The propensity score with continuous
  treatments.
\newblock \textit{Applied Bayesian modeling and causal inference from
  incomplete-data perspectives}, \textbf{22}, 73--84.

\bibitem[{Hirano et~al.(2003)Hirano, Imbens and Ridder}]{hirano2003efficient}
Hirano, K., Imbens, G.~W. and Ridder, G. (2003) Efficient estimation of average
  treatment effects using the estimated propensity score.
\newblock \textit{Econometrica}, \textbf{71}, 1161--1189.

\bibitem[{Hodges and Reich(2010)}]{hodges2010adding}
Hodges, J.~S. and Reich, B.~J. (2010) Adding spatially-correlated errors can
  mess up the fixed effect you love.
\newblock \textit{The American Statistician}, \textbf{64}, 325--334.

\bibitem[{Holland(1986)}]{holland1986statistics}
Holland, P.~W. (1986) Statistics and causal inference.
\newblock \textit{Journal of the American Statistical Association},
  \textbf{81}, 945--960.

\bibitem[{Hudgens and Halloran(2008)}]{hudgens2008toward}
Hudgens, M.~G. and Halloran, M.~E. (2008) Toward causal inference with
  interference.
\newblock \textit{Journal of the American Statistical Association},
  \textbf{103}, 832--842.

\bibitem[{Hughes and Haran(2013)}]{hughes2013dimension}
Hughes, J. and Haran, M. (2013) Dimension reduction and alleviation of
  confounding for spatial generalized linear mixed models.
\newblock \textit{Journal of the Royal Statistical Society: Series B
  (Statistical Methodology)}, \textbf{75}, 139--159.

\bibitem[{Imbens and Angrist(1994)}]{Imbens1994Identification}
Imbens, G.~W. and Angrist, J.~D. (1994) Identification and estimation of local
  average treatment effects.
\newblock \textit{Econometrica}, \textbf{62}, 467--475.

\bibitem[{Imbens and Lemieux(2008)}]{imbens2008regression}
Imbens, G.~W. and Lemieux, T. (2008) {Regression discontinuity designs: A guide
  to practice}.
\newblock \textit{Journal of Econometrics}, \textbf{142}, 615--635.

\bibitem[{Janes et~al.(2007)Janes, Dominici and Zeger}]{janes2007trends}
Janes, H., Dominici, F. and Zeger, S.~L. (2007) Trends in air pollution and
  mortality: an approach to the assessment of unmeasured confounding.
\newblock \textit{Epidemiology}, 416--423.

\bibitem[{Jarner et~al.(2002)Jarner, Diggle and
  Chetwynd}]{jarner2002estimation}
Jarner, M.~F., Diggle, P. and Chetwynd, A.~G. (2002) Estimation of spatial
  variation in risk using matched case-control data.
\newblock \textit{Biometrical Journal: Journal of Mathematical Methods in
  Biosciences}, \textbf{44}, 936--945.

\bibitem[{Keele and Titiunik(2015)}]{keele2015geographic}
Keele, L.~J. and Titiunik, R. (2015) Geographic boundaries as regression
  discontinuities.
\newblock \textit{Political Analysis}, \textbf{23}, 127--155.

\bibitem[{Kim et~al.(2018)Kim, Paini and Jurdak}]{kim2018causal}
Kim, M., Paini, D. and Jurdak, R. (2018) Causal inference in disease spread
  across a heterogeneous social system.
\newblock \textit{arXiv preprint arXiv:1801.08133}.

\bibitem[{Kim et~al.(2019)Kim, Paini and Jurdak}]{kim2019modeling}
--- (2019) Modeling stochastic processes in disease spread across a
  heterogeneous social system.
\newblock \textit{Proceedings of the National Academy of Sciences},
  \textbf{116}, 401--406.

\bibitem[{Laber et~al.(2018)Laber, Meyer, Reich, Pacifici, Collazo and
  Drake}]{laber2018optimal}
Laber, E.~B., Meyer, N.~J., Reich, B.~J., Pacifici, K., Collazo, J.~A. and
  Drake, J.~M. (2018) Optimal treatment allocations in space and time for
  on-line control of an emerging infectious disease.
\newblock \textit{Journal of the Royal Statistical Society: Series C (Applied
  Statistics)}, \textbf{67}, 743--789.

\bibitem[{Larsen et~al.(2020)Larsen, Yang, Reich and
  Rappold}]{Larsen2020spatial}
Larsen, A., Yang, S., Reich, B.~J. and Rappold, A.~G. (2020) {A spatial causal
  analysis of wildland fire-contributed PM2.5 using numerical model output}.
\newblock \textit{arXiv preprint arXiv:2003.06037}.

\bibitem[{Lindgren et~al.(2011)Lindgren, Rue and
  Lindstr{\"o}m}]{lindgren2011explicit}
Lindgren, F., Rue, H. and Lindstr{\"o}m, J. (2011) {An explicit link between
  Gaussian fields and Gaussian Markov random fields: The stochastic partial
  differential equation approach}.
\newblock \textit{{Journal of the Royal Statistical Society: Series B
  (Statistical Methodology)}}, \textbf{73}, 423--498.

\bibitem[{Liu et~al.(2016)Liu, Hudgens and Becker-Dreps}]{liu2016inverse}
Liu, L., Hudgens, M.~G. and Becker-Dreps, S. (2016) On inverse
  probability-weighted estimators in the presence of interference.
\newblock \textit{Biometrika}, \textbf{103}, 829--842.

\bibitem[{Lum et~al.(2012)Lum, Gelfand et~al.}]{lum2012spatial}
Lum, K., Gelfand, A.~E. et~al. (2012) Spatial quantile multiple regression
  using the asymmetric laplace process.
\newblock \textit{Bayesian Analysis}, \textbf{7}, 235--258.

\bibitem[{Lunn et~al.(2009)Lunn, Best, Spiegelhalter, Graham and
  Neuenschwander}]{lunn2009combining}
Lunn, D., Best, N., Spiegelhalter, D., Graham, G. and Neuenschwander, B. (2009)
  {Combining MCMC with ‘sequential’PKPD modelling}.
\newblock \textit{Journal of Pharmacokinetics and Pharmacodynamics},
  \textbf{36}, 19.

\bibitem[{McCandless et~al.(2010)McCandless, Douglas, Evans and
  Smeeth}]{mccandless2010cutting}
McCandless, L.~C., Douglas, I.~J., Evans, S.~J. and Smeeth, L. (2010) {Cutting
  feedback in Bayesian regression adjustment for the propensity score}.
\newblock \textit{The International Journal of Biostatistics}, \textbf{6}.

\bibitem[{Paciorek(2010)}]{paciorek2010importance}
Paciorek, C.~J. (2010) The importance of scale for spatial-confounding bias and
  precision of spatial regression estimators.
\newblock \textit{Statistical Science}, \textbf{25}, 107--125.

\bibitem[{Papadogeorgou et~al.(2018)Papadogeorgou, Choirat and
  Zigler}]{papadogeorgou2018adjusting}
Papadogeorgou, G., Choirat, C. and Zigler, C.~M. (2018) Adjusting for
  unmeasured spatial confounding with distance adjusted propensity score
  matching.
\newblock \textit{Biostatistics}, \textbf{20}, 256--272.

\bibitem[{Papadogeorgou et~al.(2019)Papadogeorgou, Mealli and
  Zigler}]{papadogeorgou2019causal}
Papadogeorgou, G., Mealli, F. and Zigler, C.~M. (2019) Causal inference with
  interfering units for cluster and population level treatment allocation
  programs.
\newblock \textit{Biometrics}, \textbf{75}, 778--787.

\bibitem[{Pati et~al.(2011)Pati, Reich and Dunson}]{pati2011bayesian}
Pati, D., Reich, B.~J. and Dunson, D.~B. (2011) Bayesian geostatistical
  modelling with informative sampling locations.
\newblock \textit{Biometrika}, \textbf{98}, 35--48.

\bibitem[{Perez-Heydrich et~al.(2014)Perez-Heydrich, Hudgens, Halloran,
  Clemens, Ali and Emch}]{perez2014assessing}
Perez-Heydrich, C., Hudgens, M.~G., Halloran, M.~E., Clemens, J.~D., Ali, M.
  and Emch, M.~E. (2014) Assessing effects of cholera vaccination in the
  presence of interference.
\newblock \textit{Biometrics}, \textbf{70}, 731--741.

\bibitem[{Reich(2012)}]{reich2012spatiotemporal}
Reich, B.~J. (2012) Spatiotemporal quantile regression for detecting
  distributional changes in environmental processes.
\newblock \textit{Journal of the Royal Statistical Society: Series C (Applied
  Statistics)}, \textbf{61}, 535--553.

\bibitem[{Reich and Fuentes(2015)}]{reich2015spatial}
Reich, B.~J. and Fuentes, M. (2015) {Spatial Bayesian nonparametric methods}.
\newblock In \textit{Nonparametric Bayesian Inference in Biostatistics},
  347--357. Springer.

\bibitem[{Reich et~al.(2011)Reich, Fuentes and Dunson}]{reich2011bayesian}
Reich, B.~J., Fuentes, M. and Dunson, D.~B. (2011) Bayesian spatial quantile
  regression.
\newblock \textit{Journal of the American Statistical Association},
  \textbf{106}, 6--20.

\bibitem[{Reich et~al.(2007)Reich, Hodges and Carlin}]{reich2007spatial}
Reich, B.~J., Hodges, J.~S. and Carlin, B.~P. (2007) Spatial analyses of
  periodontal data using conditionally autoregressive priors having two classes
  of neighbor relations.
\newblock \textit{Journal of the American Statistical Association},
  \textbf{102}, 44--55.

\bibitem[{Reich et~al.(2006)Reich, Hodges and Zadnik}]{reich2006effects}
Reich, B.~J., Hodges, J.~S. and Zadnik, V. (2006) Effects of residual smoothing
  on the posterior of the fixed effects in disease-mapping models.
\newblock \textit{Biometrics}, \textbf{62}, 1197--1206.

\bibitem[{Robins and Greenland(1994)}]{robins1994adjusting}
Robins, J.~M. and Greenland, S. (1994) Adjusting for differential rates of
  prophylaxis therapy for {PCP} in high-versus low-dose {AZT} treatment arms in
  an {AIDS} randomized trial.
\newblock \textit{Journal of the American Statistical Association},
  \textbf{89}, 737--749.

\bibitem[{Robins et~al.(1994)Robins, Rotnitzky and Zhao}]{robins1994estimation}
Robins, J.~M., Rotnitzky, A. and Zhao, L.~P. (1994) Estimation of regression
  coefficients when some regressors are not always observed.
\newblock \textit{Journal of the American Statistical Association},
  \textbf{89}, 846--866.

\bibitem[{Rosenbaum(1989)}]{rosenbaum1989optimal}
Rosenbaum, P.~R. (1989) Optimal matching for observational studies.
\newblock \textit{Journal of the American Statistical Association},
  \textbf{84}, 1024--1032.

\bibitem[{Rosenbaum and Rubin(1983)}]{rosenbaum1983assessing}
Rosenbaum, P.~R. and Rubin, D.~B. (1983) Assessing sensitivity to an unobserved
  binary covariate in an observational study with binary outcome.
\newblock \textit{J. R. Stat. Soc. Ser. B.}, \textbf{45}, 212--218.

\bibitem[{Rosenbaum and Rubin(1983a)}]{rosenbaum&rubin83a}
--- (1983a) The central role of the propensity score in observational studies
  for causal effects.
\newblock \textit{Biometrika}, \textbf{70}, 41--55.

\bibitem[{Rubin(1974)}]{rubin1974estimating}
Rubin, D.~B. (1974) Estimating causal effects of treatments in randomized and
  nonrandomized studies.
\newblock \textit{J Educational Psychology}, \textbf{66}, 688--701.

\bibitem[{Rubin(1978)}]{rubin1978bayesian}
--- (1978) Bayesian inference for causal effects: The role of randomization.
\newblock \textit{Ann. Statist.}, \textbf{6}, 34--58.

\bibitem[{Rubin(2006)}]{rubin2006matched}
--- (2006) \textit{{Matched Sampling for Causal Effects}}.
\newblock Cambridge, England: Cambridge University Press.

\bibitem[{Saarela et~al.(2016)Saarela, Belzile and
  Stephens}]{saarela2016bayesian}
Saarela, O., Belzile, L.~R. and Stephens, D.~A. (2016) {A Bayesian view of
  doubly robust causal inference}.
\newblock \textit{Biometrika}, \textbf{103}, 667--681.

\bibitem[{Saarela et~al.(2015)Saarela, Stephens, Moodie and
  Klein}]{saarela2015bayesian}
Saarela, O., Stephens, D.~A., Moodie, E.~E. and Klein, M.~B. (2015) {On
  Bayesian estimation of marginal structural models}.
\newblock \textit{Biometrics}, \textbf{71}, 279--288.

\bibitem[{Schnell and Papadogeorgou(2019)}]{schnell2019mitigating}
Schnell, P. and Papadogeorgou, G. (2019) Mitigating unobserved spatial
  confounding bias with mixed models.
\newblock \textit{arXiv preprint arXiv:1907.12150}.

\bibitem[{Schutte and Donnay(2014)}]{schutte2014matched}
Schutte, S. and Donnay, K. (2014) Matched wake analysis: finding causal
  relationships in spatiotemporal event data.
\newblock \textit{Political Geography}, \textbf{41}, 1--10.

\bibitem[{Sobel(2006)}]{sobel2006randomized}
Sobel, M.~E. (2006) What do randomized studies of housing mobility demonstrate?
  causal inference in the face of interference.
\newblock \textit{Journal of the American Statistical Association},
  \textbf{101}, 1398--1407.

\bibitem[{Stuart(2010)}]{stuart2010matching}
Stuart, E.~A. (2010) Matching methods for causal inference: A review and a look
  forward.
\newblock \textit{Statistical Science}, \textbf{25}, 1--21.

\bibitem[{Tchetgen et~al.(2017)Tchetgen, Fulcher and
  Shpitser}]{tchetgen2017auto}
Tchetgen, E. J.~T., Fulcher, I. and Shpitser, I. (2017) Auto-g-computation of
  causal effects on a network.
\newblock \textit{arXiv preprint arXiv:1709.01577}.

\bibitem[{Tchetgen and VanderWeele(2012)}]{tchetgen2012causal}
Tchetgen, E. J.~T. and VanderWeele, T.~J. (2012) On causal inference in the
  presence of interference.
\newblock \textit{Statistical Methods in Medical Research}, \textbf{21},
  55--75.

\bibitem[{Thaden and Kneib(2018)}]{thaden2018structural}
Thaden, H. and Kneib, T. (2018) Structural equation models for dealing with
  spatial confounding.
\newblock \textit{The American Statistician}, \textbf{72}, 239--252.

\bibitem[{VanderWeele et~al.(2014)VanderWeele, Tchetgen and
  Halloran}]{vanderweele2014interference}
VanderWeele, T.~J., Tchetgen, E. J.~T. and Halloran, M.~E. (2014) Interference
  and sensitivity analysis.
\newblock \textit{Statistical science: A Review Journal of the Institute of
  Mathematical Statistics}, \textbf{29}, 687.

\bibitem[{Verbitsky-Savitz and Raudenbush(2012)}]{verbitsky2012causal}
Verbitsky-Savitz, N. and Raudenbush, S.~W. (2012) {Causal inference under
  interference in spatial settings: A case study evaluating community policing
  program in Chicago}.
\newblock \textit{Epidemiologic Methods}, \textbf{1}, 107--130.

\bibitem[{Wall(2004)}]{wall2004close}
Wall, M.~M. (2004) {A close look at the spatial structure implied by the CAR
  and SAR models}.
\newblock \textit{{Journal of Statistical Planning and Inference}},
  \textbf{121}, 311--324.

\bibitem[{White and Lu(2010)}]{white2010granger}
White, H. and Lu, X. (2010) Granger causality and dynamic structural systems.
\newblock \textit{Journal of Financial Econometrics}, \textbf{8}, 193--243.

\bibitem[{Wu et~al.(2020)Wu, Nethery, Sabath, Braun and
  Dominici}]{wu2020exposure}
Wu, X., Nethery, R.~C., Sabath, B.~M., Braun, D. and Dominici, F. (2020)
  {Exposure to air pollution and COVID-19 mortality in the United States}.
\newblock \textit{medRxiv}.

\bibitem[{Zigler(2016)}]{zigler2016central}
Zigler, C.~M. (2016) {The central role of Bayes’ Theorem for joint estimation
  of causal effects and propensity scores}.
\newblock \textit{The American Statistician}, \textbf{70}, 47--54.

\bibitem[{Zigler et~al.(2012)Zigler, Dominici and Wang}]{zigler2012estimating}
Zigler, C.~M., Dominici, F. and Wang, Y. (2012) Estimating causal effects of
  air quality regulations using principal stratification for spatially
  correlated multivariate intermediate outcomes.
\newblock \textit{Biostatistics}, \textbf{13}, 289--302.

\bibitem[{Zigler and Papadogeorgou(2018)}]{zigler2018bipartite}
Zigler, C.~M. and Papadogeorgou, G. (2018) Bipartite causal inference with
  interference.
\newblock \textit{arXiv preprint arXiv:1807.08660}.

\bibitem[{Zigler et~al.(2013)Zigler, Watts, Yeh, Wang, Coull and
  Dominici}]{zigler2013model}
Zigler, C.~M., Watts, K., Yeh, R.~W., Wang, Y., Coull, B.~A. and Dominici, F.
  (2013) {Model feedback in Bayesian propensity score estimation}.
\newblock \textit{Biometrics}, \textbf{69}, 263--273.

\end{thebibliography}
\end{singlespace}

\section*{Appendix A.1}
Consider the true data-generating model $\bY|\bA,\bU\sim\mbox{Normal}(\beta\bA+\bU,\tau^2\bI_n)$, $\bU|\bA\sim\mbox{Normal}(\phi\bA,\bSigma_1)$ and $\bA\sim\mbox{Normal}(0,\bSigma_2)$. In this model the treatment variable and spatial process are correlated unless $\phi=0$.   If the assumed model is $\bY|\bA,\bU\sim\mbox{Normal}(\beta\bA+\bU,\tau^2\bI_n)$ and $\bU|\bA\sim\mbox{Normal}(0,\bOmega)$, or equivalently $\bY|\bA\sim\mbox{Normal}(\beta\bA,\bSigma)$ where $\bSigma=\tau^2\bI_n+\bOmega$, then the generalized least squares (and posterior mean under flat prior) estimator is ${\hat \beta}(\bA,\bY) = (\bA^T\bSigma^{-1}\bA)^{-1}\bA^T\bSigma^{-1}\bY$.  The expected value of this estimator under the true data-generating model is $\beta+\phi$ for any assumed covariance model $\bSigma$, including the model that excludes $\bU$ by setting $\bOmega=0$. 

\section*{Appendix A.2: CAR and SAR covariance models}

In Section \ref{s:confound}, we define the CAR and SAR models for individual observations, and in this section we provide the induced joint distribution of the spatial process at all $N$ locations. If $\bU\sim\mbox{CAR}(\rho,\sigma)$ then the joint distribution of  $\bU$ defined by the full conditional distributions given in Section \ref{s:confound:review} is multivariate normal with mean zero and covariance $\Sigma_{CAR}(\rho,\sigma)=\sigma^2(\bM-\rho\bW)^{-1}$, where $\bM$ is diagonal with the $i^{\text{th}}$ diagonal element $m_i$ (the number of regions neighboring region $i$) and $\bW$ has $(i,k)$ element equal one if regions $i$ and $k$ are adjacent and zero otherwise.   

Similarly, the SAR model in (\ref{e:SAR:uni}) can be solved for $\bY = (Y_1,...,Y_N)^T$ to show that the induced joint distribution is  \begin{equation}\label{e:SAR:joint}
\bY = \bA\beta + \bX\bgamma + \bvarepsilon \mbox{\ \ \ \ \ where \ \ \ \ \ } \bvarepsilon\sim\mbox{Normal}\left\{\bzero,\sigma^2(\bI_N-\psi\bC)^{-1}(\bI_N-\psi\bC)^{-1}\right\}
\end{equation}
with the ($i,k$) element of $\bC$ is $1/m_i$ if regions $i$ and $k$ are adjacent and 0 otherwise, so that, e.g.,  $\bC\bY=({\bar Y}_1,...,{\bar Y}_N)^T$ is the vector of neighborhood means.  
    
\section*{Appendix A.3: Details of \cite{schnell2019mitigating}}

\cite{schnell2019mitigating} provided a set of assumptions to identify the unmeasured confounding bias $\mbox{E}(U_i|\bA)$. They assume a joint distribution for ($\bU$, $\bA$) that is multivariate normal with mean zero and covariance
$$ \mbox{Cov}\begin{pmatrix} \bU\\ \bA\end{pmatrix}=\begin{pmatrix} \bQ_U & \bQ_{UA}\\ \bQ_{UA}^T & \bQ_A \end{pmatrix}^{-1},$$ 
where $Q_j = \sigma_j^{-2}(\bM-\rho_j\bW)$ for $j\in\{U,A\}$ and $\bQ_{UA} = -\rho\sigma_U\sigma_A\bM$. Two assumptions are encoded in $\bQ_{UA}$: (1) a cross-Markov relationship such that conditional on all other locations treatments $\bA_{-i}$, the local treatment $A_i$ is only correlated with the local confounder $U_i$ \citep[e.g.,][]{reich2007spatial}, and (2) the conditional correlation between $A_i$ and $U_i$ is constant in space. The confounding bias $B(\bA) = \mbox{E}(\bU|\bA)= -\bQ_U^{-1}\bQ_{UA}\bA$ is mitigated by fitting a spatial model with confounder adjustment,
\begin{eqnarray}\label{e:CAR:adjust1:App}
\bY &=& \bA\beta - B(\bA) + \bX\bgamma + \boldsymbol{e} \mbox{\ \ \ \ \ where \ \ \ \ \ } \boldsymbol{e}\sim\mbox{Normal}\left\{\bzero,\bQ_U^{-1} + \tau^2 \bI_N\right\}\nonumber\\
\bA &\sim&\mbox{Normal}\left[\bzero,\sigma_A^{2}\left\{(\bM-\rho_A\bW) -\rho^2\bM^T(\bM-\rho_U\bW)^{-1}\bM \right\}^{-1}\right].\nonumber
\end{eqnarray}

\end{document}